\documentclass[twocolumn]{aastex631}

\usepackage{cancel}
\usepackage{amsmath}
\shorttitle{MeerKAT Polarization Survey}
\shortauthors{Taylor and Legodi}
\graphicspath{{./}{figures/}}

\begin{document}

\title{A MeerKAT Polarization Survey of Southern Calibration Sources}

\author[0000-0001-9885-0676]{A.\ R.\ Taylor}
\affiliation{Inter-University Institute for Data Intensive Astronomy, and \\
Department of Astronomy\\
University of Cape Town \\
Private Bag X3, Rondebosch
Cape Town, 7701, South Africa}
\affiliation{Department of Physics and Astronomy\\
University of the Western Cape \\
Robert Sobukwe Road, Bellville \\
Cape Town, 7535 South Africa}

\author[0000-0001-5205-8501]{L.\ S.\ Legodi}
\affiliation{South African Radio Astronomy Observatory \\
2 Fir Street, Black River Park,\\
Cape Town, 7925, South Africa}



\begin{abstract}

We report on full-Stokes L-band observations of 98 MeerKAT calibration sources. 
Linear polarization is detected in 71 objects above a fractional level of 0.2\%. 
We identify ten sources with strong fractional linear polarization and low
Faraday Rotation Measure that could be suitable for wide-band absolute polarization calibration. 
We detect significant circular polarization from 24\% of the sample down to a
detection level of 0.07\%.  Circularly polarized emission is seen only for flat spectrum sources $\alpha > -0.5$.

We compare our polarized intensities and Faraday Synthesis results to data
from the NVSS at 1400\,MHz and the ATCA SPASS survey at 2300\,MHz.  NVSS 
data exists for 54 of our sources and SPASS data for 20 sources.  The percent polarization and Rotation Measures from both surveys agree well with our results.

The residual instrumental linear polarization for these observations is measured at 0.16\% and
the residual instrumental circular polarization is measured at 0.05\%.  
These levels may reflect either instabilities in the relative bandpass between the two
polarization channels with either time or antenna orientation, or atmospheric/ionospheric 
variations with pointing direction. Tracking of the hourly gain solutions on J0408-6545 after
transfer of the primary gain solutions suggests a deterioration of the gain stability by a factor
of several starting about two hours after sunrise.  This suggests that observing during the nighttime
could dramatically improve the precision of polarization calibration.

\end{abstract}



\section{Introduction} \label{sec:intro}

Radio polarimetry is a powerful probe of cosmic magnetic fields, and is central to science programs on next generation radio facilities and to planning for the Square Kilometre Array \citep{Heald_2020}.
For telescopes with linear polarized feeds, polarized signals from
radio sources are mixed into all correlation products. 
It is therefore important for both observational planning and precise
calibration to have knowledge of the polarization of
calibration sources.
To this end, we have begun a program to observed the polarization properties of MeerKAT 
gain calibration sources.  In this paper we report on polarization observations of 
an initial set of 98 MeerKAT gain calibration sources at L-band.

These calibrators were chosen from a list of 448 sources with declinations below +34 degrees. This initial list was made up of sources from the master list of \citep{Dixon1970ApJS}, PKSCAT90 \citep{Wright1990PKSCAT90}, NVSS \citep{Condon199NVSS}, SUMSS v 2.1 \citep{Mauch2008SUMSS2.1} with additional data from the The Parkes-MIT-NRAO (PNM) surveys \citep{Griffith1993PNM1,Condon1993PNM4,Wright1994PNM2,Griffith1994PNM3,Tasker1994PNM5} and SPECFIND \citep{Vollmer2005a,Vollmer2005Cat,Vollmer2010}. All these are available on the VizieR database. The list also includes the well known VLA calibrators 3C48 (J0137+3309), 3C138 (J0521+1638), 3C286 (J1331+3030) as well as 
PKS B1934-638 (J1939-6342), which is the main calibrator for the Australia Telescope National Facility (ATNF). Another incorporated list was made from VLA calibrators with flux densities greater than 2 Jy at L-band and declinations below -20 degrees. Some of the Molonglo Observatory calibrators were also added to the candidate list but these sources did not pass the final criteria. The criteria for the final list being that sources must: 1) have L-band flux densities above 1 Jy, and be stronger than 10 times the background confusion; 2) be no more than $10\%$ resolved; 3) have closure phases less than 5 degrees ($\sim 0.1$ radians); 4) have stable flux densities on year timescales.


\section{Observations}
\label{sec:obs}
The calibration sources were observed with MeerKAT \citep{Jonas_2016} during one of two observing runs carried out by the SARAO MeerKAT commissioning team. MeerKAT receivers have two orthogonal linear feeds and the correlator produces all four polarization products viz. HH, HV, VH and VV\footnote{\url{https://skaafrica.atlassian.net/wiki/spaces/ESDKB/pages/1493631000/Polarisation+calibration}} during all observing runs.  The first observing run occurred on 19 August 2019 and the second run on 29 August 2020.  The first run observed 52 sources in the RA range 11--23 hours, and second run observed 46 sources in the RA range 00-11 hours.  Both observing runs were about 12 hours long. Each target source was observed for a short scan of about 5 minutes. The strong unpolarized source J1939-6342 was observed several times during the each run, and the absolute polarization calibrator J1331+3030 (3C286) was observed twice for 5 minutes at the beginning of the first run, and once for 10 minutes at the end of the second run.

\section{Calibration and Imaging}
\label{sec:cal}
The data were processed on the Ilifu cloud using a customized version of the IDIA calibration and imaging pipeline\footnote{\url{https://idia-pipelines.github.io/docs/processMeerKAT}} \citep{pipeline}. The calibration stage of the pipeline utilises CASA processing \citep{casa2022}.  The pipeline partitions the L-band RF into 15 equally-sized sub-spectral windows between 880\,MHz to 1680\,MHz.  Data in frequency ranges with strong persistent RFI (933--960\.MHz, 1163--1299\,MHz, and 1525--1630\,MHz) are removed at this partition stage.  Each spectral window 
is split into its own multi-MS that is processed concurrently using the SLURM job manager.  Following calibration the calibrated data from each spectral
window is merged into a single measurement set with calibrated visibilities.

For linear feeds, the linear approximation to the response of the parallel and cross-hand visibilities is given in equations 1-4 below.  
In these equations we have retained only those leakage terms that multiply Stokes $I$. A more complete set including leakage terms that multiply the source polarization can be
found in \cite{Hales_2017}.
{\small
\begin{eqnarray}
V_{xx} = g^i_{x}g^k_{x} \bigl(I + Q\cos2\psi + U\sin 2\psi \bigr) \\
V_{xy} = g^i_{x}g^k_{y}\bigl [ (d^i_x - d^k_y\ ^*)I - Q \sin 2\psi + U\cos 2\psi + jV\bigr ] \\
V_{yx} = g^i_{y}g^k_{x}\bigl [ (d^k_x\ ^* - d^i_y)I - Q \sin 2\psi + U\cos 2\psi - jV \bigr ] \\
V_{yy} = g^i_{y}g^k_{y} \bigl(I - Q\cos2\psi - U\sin 2\psi \bigr) 
\end{eqnarray}
}
Any polarized signal from the source is present in all four correlations and varies with parallactic angle $\psi$. 
Precise measurement of the gains and the leakage terms can be 
achieved using a strong unpolarized source for which the equations reduce to 
\begin{eqnarray}
V_{xx} = g^i_{x}g^k_{x} I \\
V_{xy} = g^i_{x}g^k_{y} (d^i_x - d^k_y\ ^*)I  \\
V_{yx} = g^i_{y}g^k_{x} (d^k_x\ ^* - d^i_y)I  \\
V_{yy} = g^i_{y}g^k_{y} I
\end{eqnarray}
These equations cleanly separate the gain solution into the parallel-hand correlations and the leakage into 
the cross-hand correlations. As a first step in the calibration we use the observations of J1939-6342 
to measure the frequency-dependent $g_x$ and $g_y$ gains.   We then apply the gain solutions and derive
the leakage terms also using J1939-6342.

 All of the target sources are strong compact objects at the phase and pointing centre of the observation. 
 Following the application of gain and leakage solutions from J1939-6342 to each source, gain solutions for each source are derived by running CASA's \texttt{gaincal} task on each source (essentially a self-calibration assuming a point source model).  
 We use \texttt{gaincal} parameter \texttt{gaintype = `T'}, which averages the $xx$ and $yy$ data before solving for the gain.  The local residual gain solution for each source therefore does not modify the relative values of $g_x$ and $g_y$ gains derived from the bandpass calibrator, thus preserving the polarization calibration, and the residual gain solution is not affected by any polarization of the source, as can be seen from inspection of equations 1 and 4.  
 \begin{equation}
     V_{xx}+V_{yy} = (g^i_{x}g^k_{x}  + g^i_{y}g^k_{y}) I
 \end{equation}
 The residual gain solution for each source is thus a local, polarization-independent correction to the absolute gain solution from J1939-6342.
 
\begin{figure*}
    \centering
    \includegraphics[width=0.8\textwidth]{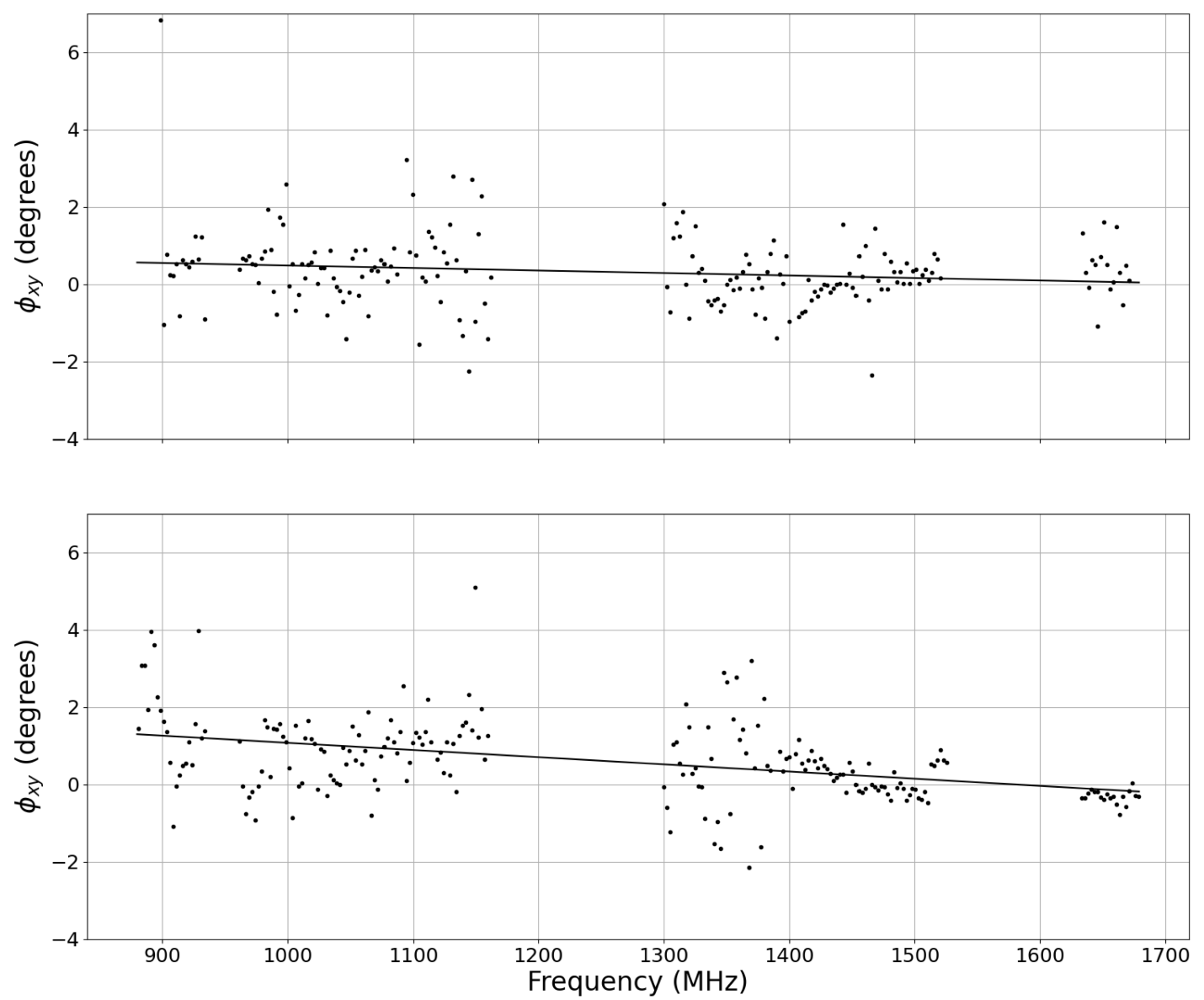}
    \caption{The residual $x-y$ phase angle (degrees) measured from the Stokes $U$ and $V$ spectra of J1331+3030 (3C\,286). The top panel shows $\phi_{xy}$ versus frequency for the first run and the bottom for the second run.  The lines are a linear fit.}
    \label{fig:xyphase}
\end{figure*} 

The target sources are distributed over the sky -- as far North as +33 and as far South as -84 declination.  The transfer of the absolute gain solution from J1939-6342 thus occurs across a large angle on the sky.  At GHz frequencies the gain of the system is largely determined by the instrument.  The atmosphere plays a minor role.  This observing and calibration approach is thus practical for frequencies in the GHz regime.  At higher frequencies where the atmosphere has a strong impact on the gains, or at lower frequencies where the ionosphere will strongly affect the $x-y$ phase, this approach may not be practical.

The last step of the calibration is to solve for any remaining $x-y$ phase using the scan on the
polarization calibrator J1331+3030 (3C286). 
A residual error in $x-y$ phase has the effect of rotating flux between $U$ and $V$. For a linear polarized source known to have no circular polarization $V=0$, the $x-y$ phase residual can be measured as
\begin{equation}
    \Delta \phi_{xy} = \arctan \biggl (\frac{V}{U} \biggr )
\end{equation}
The residual $x-y$ phase on 3C\,286 as a function of frequency for each observing run is shown in Figure ~\ref{fig:xyphase}.  The residual phase is small, with a maximum of about 2$^{\circ}$ at the low end of the band and close to zero at the high end.  An $x-y$ phase error of 2$^{\circ}$ results in a rotation of 0.06\% between $U$ and $V$. While small, under the assumption that the Stokes $V$ flux of 3C\,286 is actually zero, this final residual $x-y$ phase correction (solid lines in Figure~\ref{fig:xyphase}) is 
applied to each source.

\begin{figure}[htb]
    \centering
    \includegraphics[width=\columnwidth]{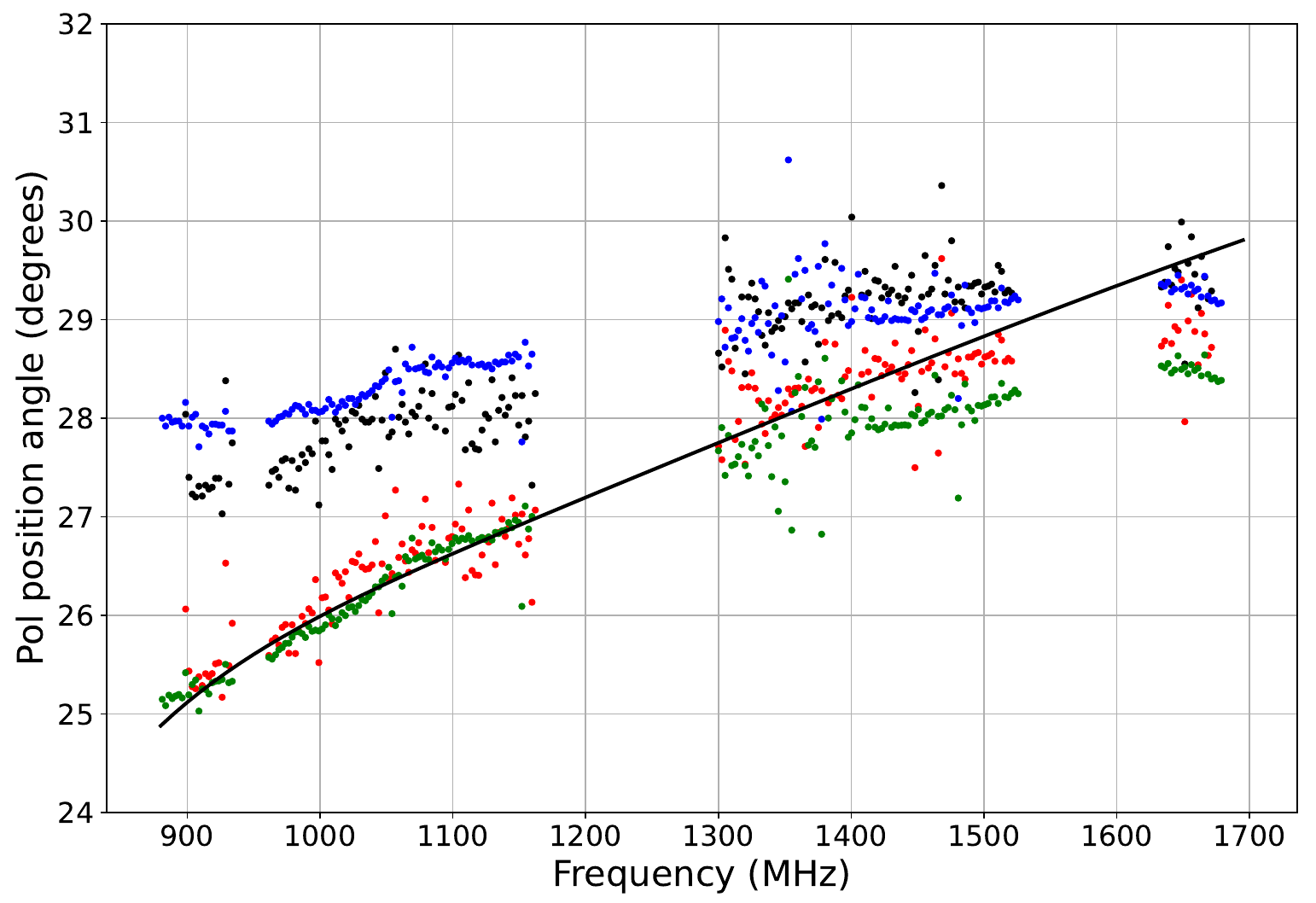}
    \caption{The observed polarization position angle versus frequency for 3C\,286 on 19 August 2019 (black dots) and 29 August 2020 (blue dots).   The solid black line
    is the revised model for the polarization position angle of 3C\,286 from \cite{Hugo_2023}.
    The red and green dots are the data from 2019 and 2020
    with an additional correction to align with the model
    (see text).}
    \label{fig:polpa}
\end{figure}

Following calibration, a full Stokes cube for each source was created using the IDIA cube generation pipeline component\footnote{\url{https://github.com/idia-astro/mightee-pol}} written by Lennart Heino. 
The cubes contain Stokes $I$,\ $Q$,\ $U$,\ $V$ images versus frequency in 320 channels, each of width 2.51\,MHz. Each channel image is $512\times512$ pixels with pixel cell size of 1.5$''$.   
Full Stokes spectra for each source were constructed by extracting the peak flux in the $I$,\ $Q$,\ $U$ and $V$ at the location of the peak Stokes $I$ flux density in each channel.
The spectral noise for each source was determined by measuring the variance of intensity with frequency for an off-source position in each spectral cube. 
The median spectral noise per-channel over all sources in $I,\ Q,\ U$ and $V$ is 2.61, 0.30, 0.32 and 0.23 mJy-bm$^{-1}$ respectively.
Spatial noise per-channel was measured from the standard deviation in a region around the source in each spectral channel.

As a final step, $Q,\ U$ spectra for each source were corrected for
ionospheric Faraday rotation using predicted ionospheric 
Rotation Measures from the World Magnetic Model (WMM)\footnote{\url{https://www.ncei.noaa.gov/products/world-magnetic-model}}\citep{WMM2020} through the use of the \texttt{RMExtract} software package\footnote{\url{https://github.com/lofar-astron/RMextract}} \citep{Mevius2018}. \texttt{RMExtract} predicts ionospheric total electron content (TEC) values and rotation measures for a given line of sight and observation time from a geomagnetic field model and Global Navigation Satellite System (GNSS) global ionospheric map\footnote{\url{https://www.aiub.unibe.ch/research/code___analysis_center/index_eng.html}}.

The observed frequency dependence of the polarization position angle for 3C\,286 for each observing run is shown by the black and blue dots in Figure~\ref{fig:polpa}.  
The position angle is similar between runs. The difference between the observed position angle is essentially zero at the 
high end of the band, increasing to about $0.8^{\circ}$ at 900 MHz.

The solid black line in Figure~\ref{fig:polpa} shows a \textbf{recent} model of the intrinsic polarization position angle of
3C\,286 from \cite{Hugo_2023} using  
MeerKAT observations of the Moon to calibrate
absolute position angle. 
\textbf{The model agrees with the observed angles above
1500\,MHz, but lies below the observations by a few
degrees at lower frequencies.}
The red and green data points in Figure~\ref{fig:polpa} are the observed results from the two observing runs adjusted for a RM of -0.31 rad\,m$^{-2}$ for 19 August 2019 and -0.4 rad\,m$^{-2}$ for 29 August 2020 to align
the position angles with the \cite{Hugo_2023} model.   The RM adjusted data agree well with the predicted intrinsic
values below 1400\,MHz, and differ only by about
1$^{\circ}$ at the high end of the band.
These results indicate that residuals to the \textbf{ predicted
ionospheric Faraday rotation measures from RMextract} are of order -0.4 rad\,m$^{-2}$, resulting in systematic
errors in the polarization position angle at 1400\,MHz
of $\sim$2$^{\circ}$.

\subsection{Residual Instrumental Polarization}
\label{sec:leakage}

The median noise in fractional polarization $Q/I,\ U/I$ and $V/I$ is 0.011, 0.011 and 0.008\%, respectively.
The limits to the precision of polarization measurements in our observations are higher than this due to residual instrumental polarization.
J1939-6342 is assumed to have zero linear and circular polarization.
We note that  \cite{Rayner_2000} have reported a tentative detection of circular
polarization of J1939-6342 at 4.8\,GHz of $+0.029\pm0.005\%$, so 
there may be very low levels of source signal present in $V$.

\begin{figure*}
    \centering
    \includegraphics[width=0.8\textwidth]{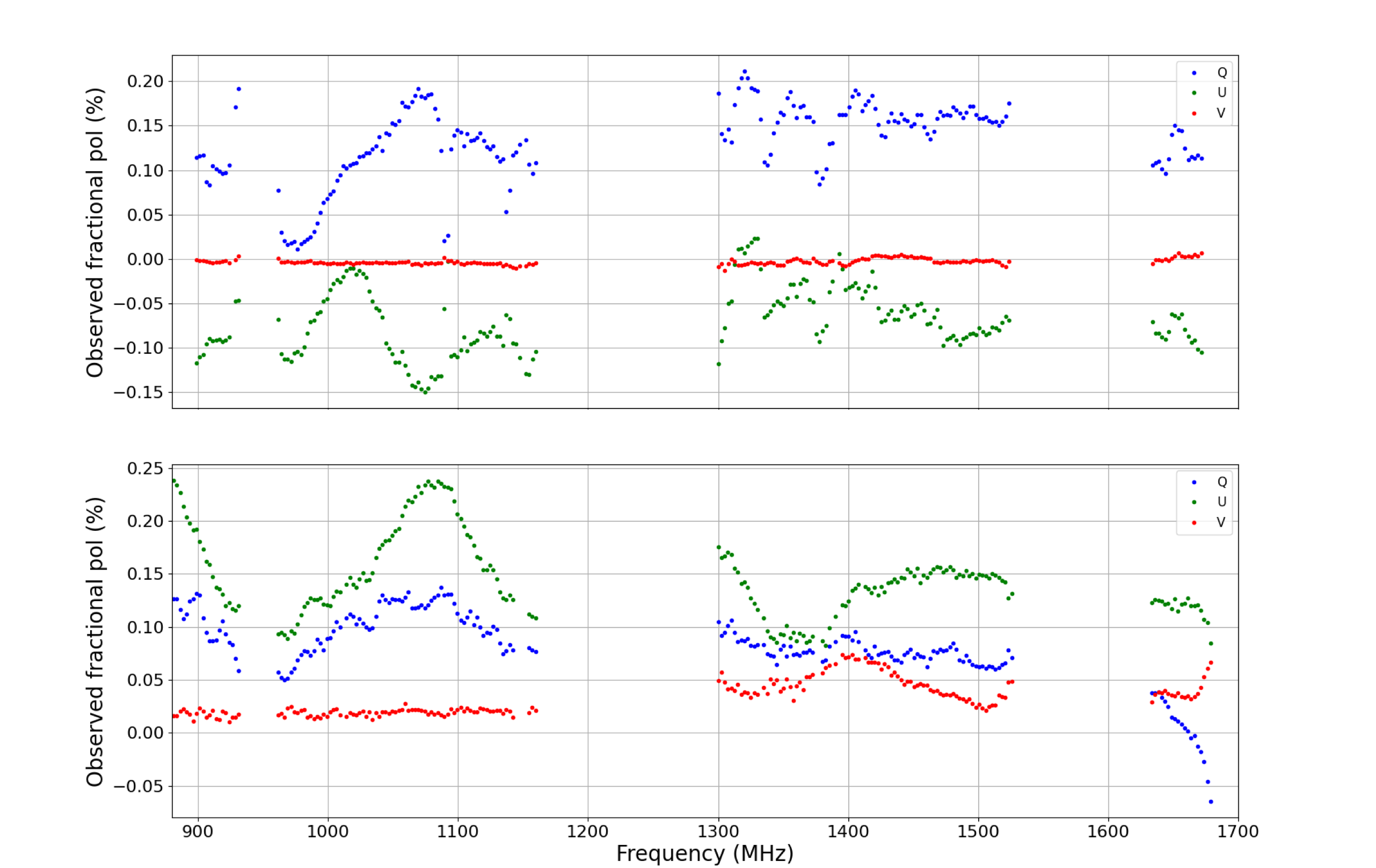}
    \caption{Fractional polarization spectra of J1939-6342 on 19 August 2019 (top) 
    and 29 August 2020 (bottom).  For each run, traces are shown for 
    $Q/I$ (blue),\ $U/I$ (green) and $V/I$ (red).
    }
    \label{fig:J1939_residuals}
\end{figure*}

Since J1939-6342 was observed several times for
each observing run, and is the primary polarization 
leakage calibrator, the residual polarized signal on this
sources represents an estimate of the residual 
error of the fractional polarization.
Figure~\ref{fig:J1939_residuals} shows the spectrum of polarized signals for J1939-6342 for the two observing runs.  
The polarization spectra shows signal with structure in frequency and of about 0.1\% in linear polarization - about a factor of 10 higher than the RMS noise. The peak-to-peak variation of the signal in $Q/I$ and $U/I$ across the band is similar to the average signal strength.
The circularly polarized signal is significantly 
lower but still present. Maximum peak absolute values in 
linear polarization are 0.25\% and 0.07\% in
circular polarization.
The median and rms values over the band for
each Stokes parameter are listed in Table~\ref{tab:J1939}.
Median linear polarization of J1939-6342 over the band is $\sim0.15$\% for both observing runs.
The median circular polarization varies between runs
with a maximum value of 0.026\%.
Conservatively, we set the threshold for
polarization detection for linear polarization as band-averaged value greater than 0.2\% and median circular polarization greater than 0.08\%.

\begin{table*}
\small
    \centering
    \begin{tabular}{c|ccc|ccc}
     \hline \hline
     Date & $Q/I$ & $U/I$ & $V/I$ &  $Q/I$ &  $U/I$ &  $V/I$ \\
     && Median (\%) &&& RMS (\%) & \\
     \hline
     19 August 2019 & 0.082 & ~0.142 & ~0.026 & 0.033 & 0.039 & 0.016 \\
     29 August 2020 & 0.141 & -0.075 & -0.004 & 0.044 & 0.037 & 0.003 \\
     \hline \hline
    \end{tabular}
    \caption{Band-averaged polarization of J1939-6342 for the two observing runs.}
    \label{tab:J1939}
\end{table*}

Residual leakage signals may occur for two main reasons. Firstly, they may stem from instabilities in the bandpass, leading to time variations in the frequency spectrum of the leakage. Secondly, they could arise due to directional variations, either caused by minor changes in the optics concerning antenna orientation or external factors, such as ionospheric effects. These external influences have the potential to impact both the gain phase and the rotation angle of polarization.
As noted in section~\ref{sec:obs}, the ionosphere is expected to
introduce a polarization dependent phase variation of order a few 
degrees at L-band.

During the 19 August 2019 run J1939-6342 was observed five times over a time span of about seven hours.
For the 20 August 2020 run, J1939-6342 was observed three times over a much shorter span of just under three hours.  
For the polarization dependent bandpass solution, the solution interval
was set to 60 minutes to provide a crude time-dependent solution.
For the leakage calibration the solution interval was set to be infinite through the `inf' option of CASA's \texttt{polcal()} task, to create a 
single average leakage solution.  
Any instabilities or the direction dependence from the average value will propagate through
as residual errors.  

\begin{figure*}
    \centering
    \includegraphics[width=0.9\textwidth]{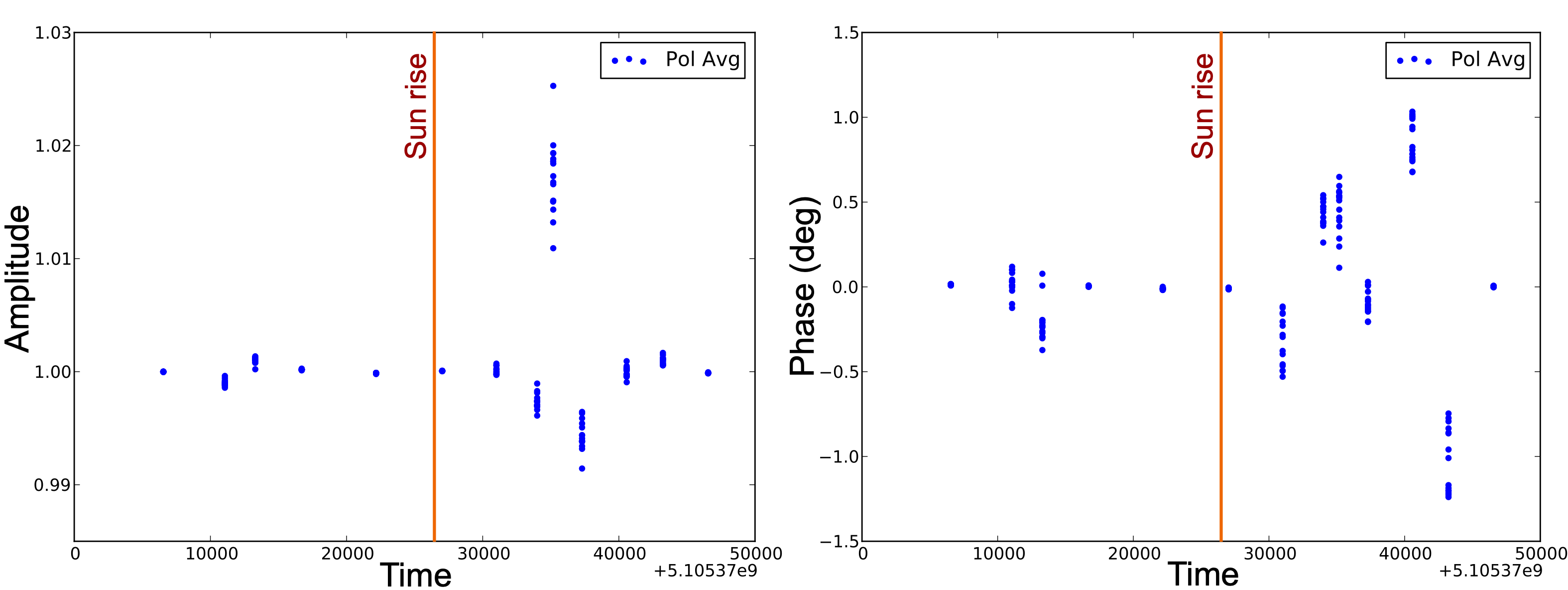}
    \caption{Gaincal solutions for hourly observations of J0408-6545 during the
    observing run of 29 August 2020. The x-axis shows observation time stamps with solution amplitude and phase given on the y-axes.  The left hand panel shows gain amplitude 
    versus time and the right hand panel the gain phase versus time.  The
    vertical red line shows the approximate time of astronomical Sunrise.
    }
    \label{fig:0408_gains}
\end{figure*}

One may not expect ionospheric effects to result in the frequency structure seen in Figure~\ref{fig:J1939_residuals}.
However, one way to test for the impact of the ionosphere is to carry out an observing run at nighttime when the 
ionosphere is not excited by solar radiance.  
An indication of the impact of nighttime versus daytime observation is shown in 
Figure~\ref{fig:0408_gains}.  For the second run in August 2020 we observed the
strong source J0408-6545 approximately once per hour during the course of the run.
The Figure shows the gain amplitude and phase solutions from \texttt{gaincal} as a function 
of time.  These gain solutions were derived after applying the gain solutions from
the primary calibrator J1939-6342 and were made with \texttt{gaintype=`T'} so the parallel
hand correlations are averaged before the solve.  If the J1939-6342 solutions
were correct for J0408-6545 then the solution gain amplitude solution would be 1.0 and the 
phase solution would be 0.0.   The vertical red line on each plot shows the approximate
time of astronomical Sunrise on the date of the observation.
It is noteworthy that prior to Sunrise the gain solutions are quite stable and uniformly
close to 1.0 in amplitude and 0.0 in phase. The peak-to-peak dispersion is approximately $\pm$0.1\% in gain and $\pm$0.3 degrees in phase.  About two hours after astronomical Sunrise the dispersion becomes significantly larger, 1-2\% in gain and over 1 degree
in phase.  This strongly suggests that the atmospheric/ionospheric stability is much 
better during the night (by a factor of several to ten) and thus transfer of gain solutions from the primary calibrator to target sources more precise.  This may reduce the
level of residual instrumental polarization.

\section{Results}

\subsection{Total Intensity Properties}
The polarization properties for each source were derived from the
spectro-polarimetric data for each source extracted from the $IQUV$ cubes. 
The in-band total intensity spectra was fit by a power law function of the form
\begin{equation}
    I(\nu) = I_o \biggl(\frac{\nu}{\nu_o}\biggr )^{\alpha + C \ln{(\nu/\nu_o)}}.
    \label{eqn:specfit}
\end{equation}
where $\nu_o = 1400$\,MHz.  The curvature term $C$ was included
since many of the sources exhibit curved spectra.   For example, 
12 of the sources exhibit GigaHerz Peak Spectra (GPS) and 7 show
evidence of a ``valley'' spectrum, with a minima of intensity
within the band.  In total, about 50\% of the sources have a significant curvature term.

\begin{figure*}
    \centering
    \includegraphics[width=0.9\textwidth]{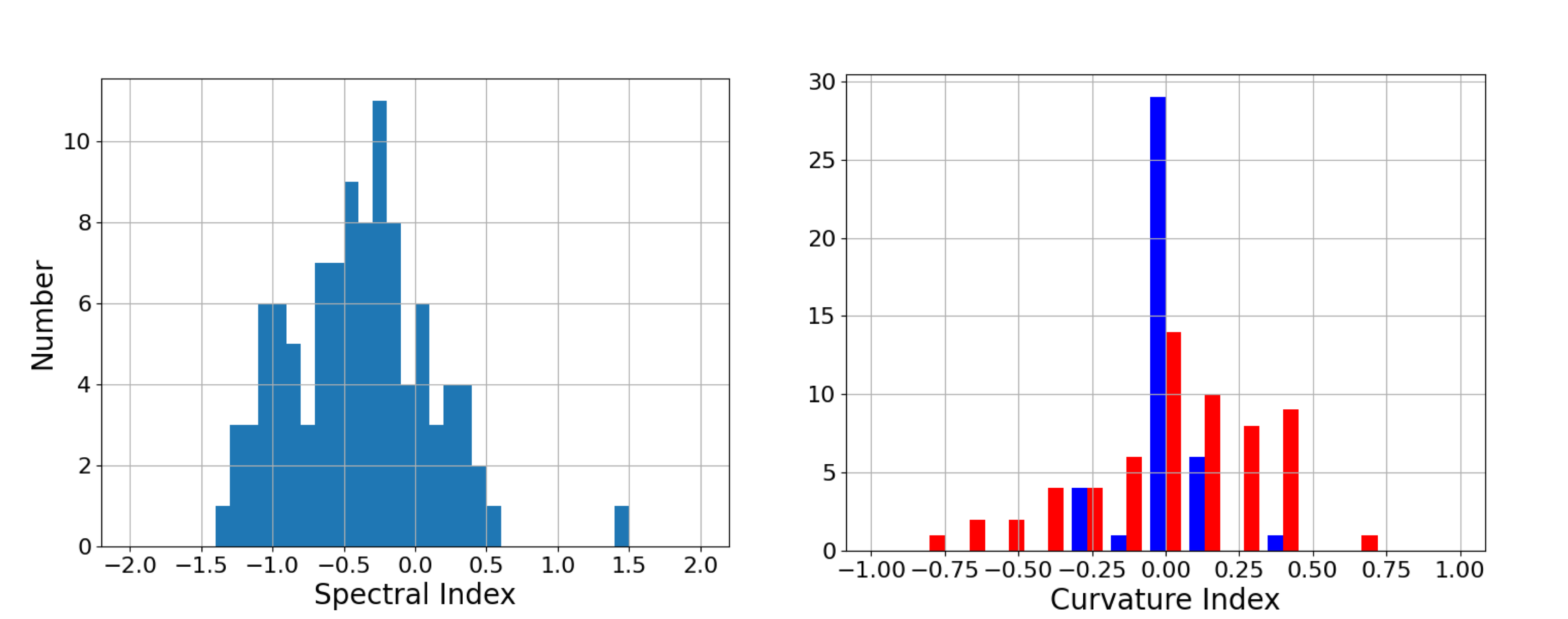}
    \caption{Left panel: The distribution of in-band spectral indices, $\alpha$. Right panel: The distribution of in-band curvature indices. The blue
    bars are for sources with spectral index $\alpha < -0.5$, and the red for $\alpha > -0.5$.}
    \label{fig:alphadist}
\end{figure*}

Figure~\ref{fig:alphadist} shows the distribution of in-band spectral and curvature indices for the sources. 
The median spectral index is $-0.39$.  This is flatter than typical of the general source population, which is more typically $-0.7$.  This difference can be attributed to the fact that the calibrator sources are
selected for brightness and compactness.  
The optical depth to synchrotron self-absorption is  proportional to
$(S^{\frac{2}{5}} \theta^{-\frac{4}{5}})$, where $\theta$ is the angular dimension of the source, so synchrotron self-absorption will be more dominant than in the general
population.

The colours in the curvature index distribution in Figure~\ref{fig:alphadist} show the values for flat spectrum
sources ($\alpha > -0.5$) in red and for steep spectrum
sources ($\alpha < -0.5)$ in blue.  Large values 
of curvature are predominantly associated with flat 
spectra.

\begin{figure}[h!]
    \centering
    \includegraphics[width=\columnwidth]{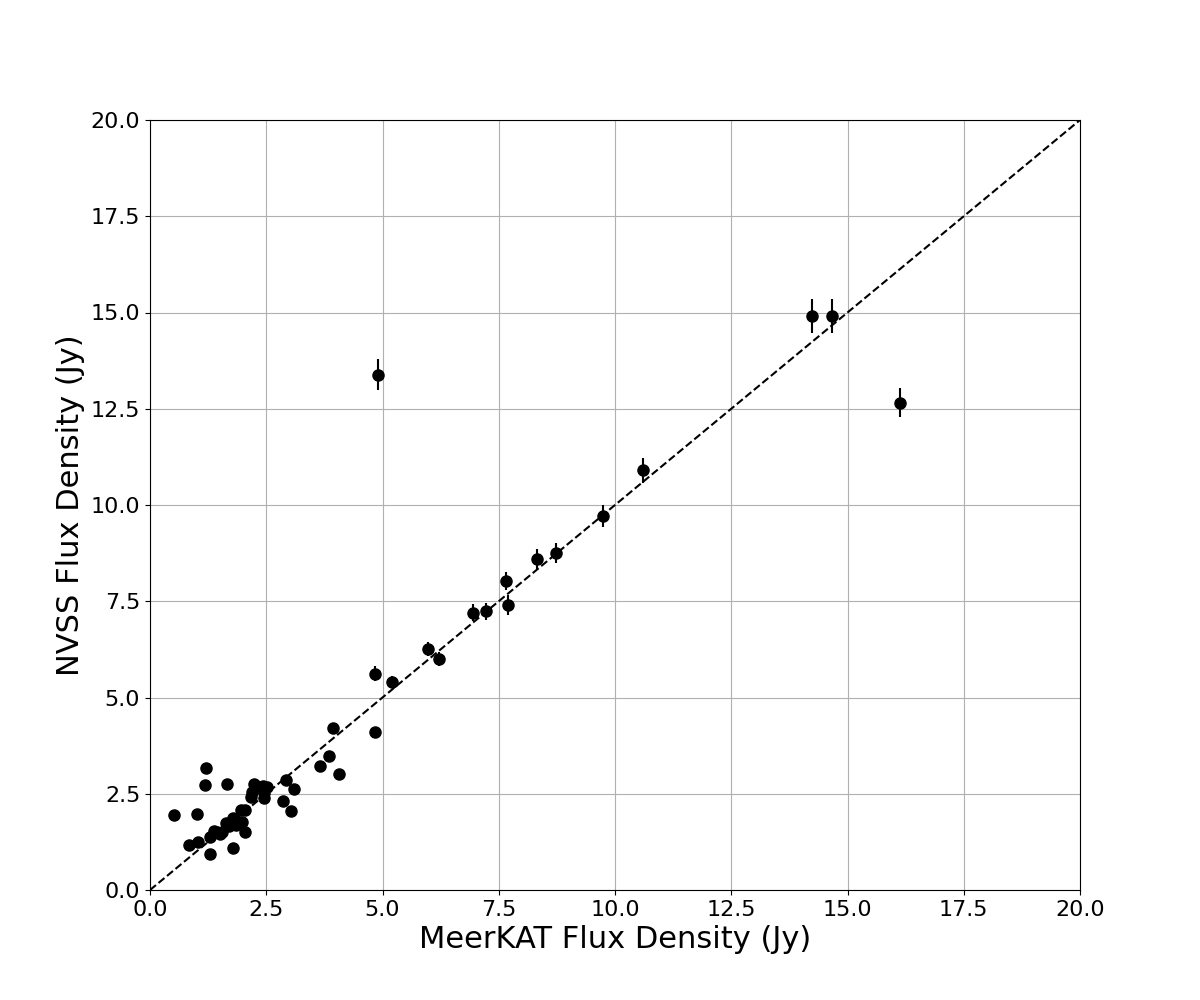}
    \caption{Comparison of the Flux density at 1400\,MHz from MeerKAT and NVSS. The dashed line
    shows the 1:1 relation.}
    \label{fig:fluxflux}
\end{figure}

Figure~\ref{fig:fluxflux} compares the flux density at 1400\,MHz from MeerKAT and NVSS.  The dashed line is not a fit. It shows the 1:1 relation. There is a core of sources that fall on the 1:1 line.  These sources are apparently flux stable on time scales of 20 years.  Two strong sources exhibit high fractional variability, J1924-2914 decreased from 13.4\,Jy to 4.9\,Jy since the mid 1990's when the NVSS observation were made.  J2253+1608 increased from 12.7\,Jy to 16.2\,Jy.  Several of the fainter sources also show 
variability.  Again, given the compactness criteria for the source selection, the presence of significant variability, particularly over time scales of decades, is not unexpected.

\subsection{Linear Polarization Properties of the Calibrators}

For each source we derive three measures of the linear polarized signal.  A band-averaged fractional linear polarization of the calibrator sources was measured by taking the 
median polarized intensity $p_{\mathrm med}$ over
all frequency channels.  To correct for polarization bias from noise, we measured
the standard deviation in $Q$ and $U$ for each channel over an annular box centred on the source position. The per-channel polarized intensity is the ratio of the bias-corrected polarized intensity to the total peak intensity
in each channel. 
\begin{equation}
    p_i = \frac{\sqrt{Q_i^2 + U_i^2 - 2.3\sigma_{\rm {QU}_i}^2}}{I_i},
\end{equation}
where $i$ is the channel number and $\sigma_{\rm {QU}_i}$ is the mean of $\sigma_{\rm Q_i}$ and $\sigma_{\rm U_i}$ for each channel.
The bias correction follows the prescription of
\cite{George_2012}.
$p_{\mathrm med}$ is the median of the $p_i$ values.
The  mean frequency of the channels is 1250.2\,MHz.
The error on $p_{\mathrm med}$ is calculated as
\begin{equation}
    \sigma_n = \frac{1}{N} \sum_i^N \biggl ( \frac{\sigma_{QU_i}}{I_i} \biggr )\frac{1}{\sqrt{N}}
    \label{eqn:p_err}
\end{equation}
Here $N$ is the number of channels. 
This is the error from the noise, and does not
include the systematic error. 

For comparison to other published results, e.g.\ the NVSS, we also calculate $p_{\mathrm 1400}$, the polarized intensity at a frequency of 1400\,MHz, as the average polarized intensity over all channels within $\pm$20\,MHz of 1400\,MHz.
The error is given by Equation~\ref{eqn:p_err}, with
$N$ in this case given by the number of channels used
for the calculation (typically 15).
The linear polarization position angle
at 1400\,MHz is calculated from the average $Q_{\mathrm 1400}$ and $U_{\mathrm 1400}$. 

For a third measure of the polarized intensity we calculate a Faraday depth spectrum for
each source and record the value of the peak of the dominant component, $p_{\mathrm max}$. For
a Faraday simple spectrum (a single Faraday-thin component) the $p_{\mathrm max}$ will be similar to $p_{\mathrm med}$.  
In the presence of multiple Faraday components $p_{\mathrm max}$ will underestimate the 
total polarized intensity of the source.

\begin{figure}[h]
    \centering
    \includegraphics[width=\columnwidth]{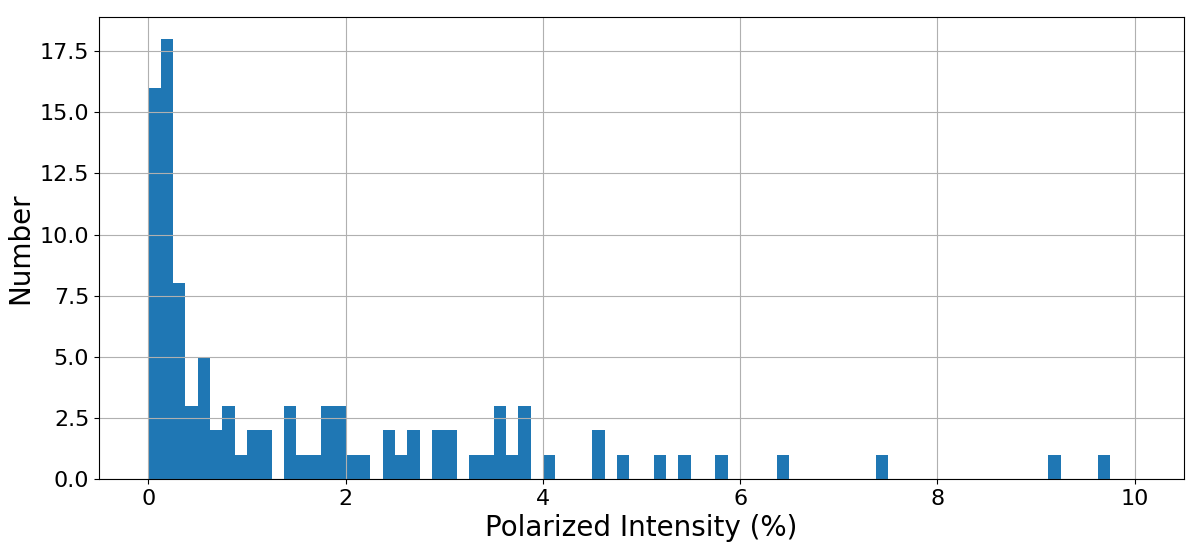}
    \caption{Distribution of linear fractional polarization $p_{\mathrm med}$ for the calibrator sources.}
    \label{fig:pdist}
\end{figure}

\begin{figure}[]
    \centering
    \includegraphics[width=\columnwidth]{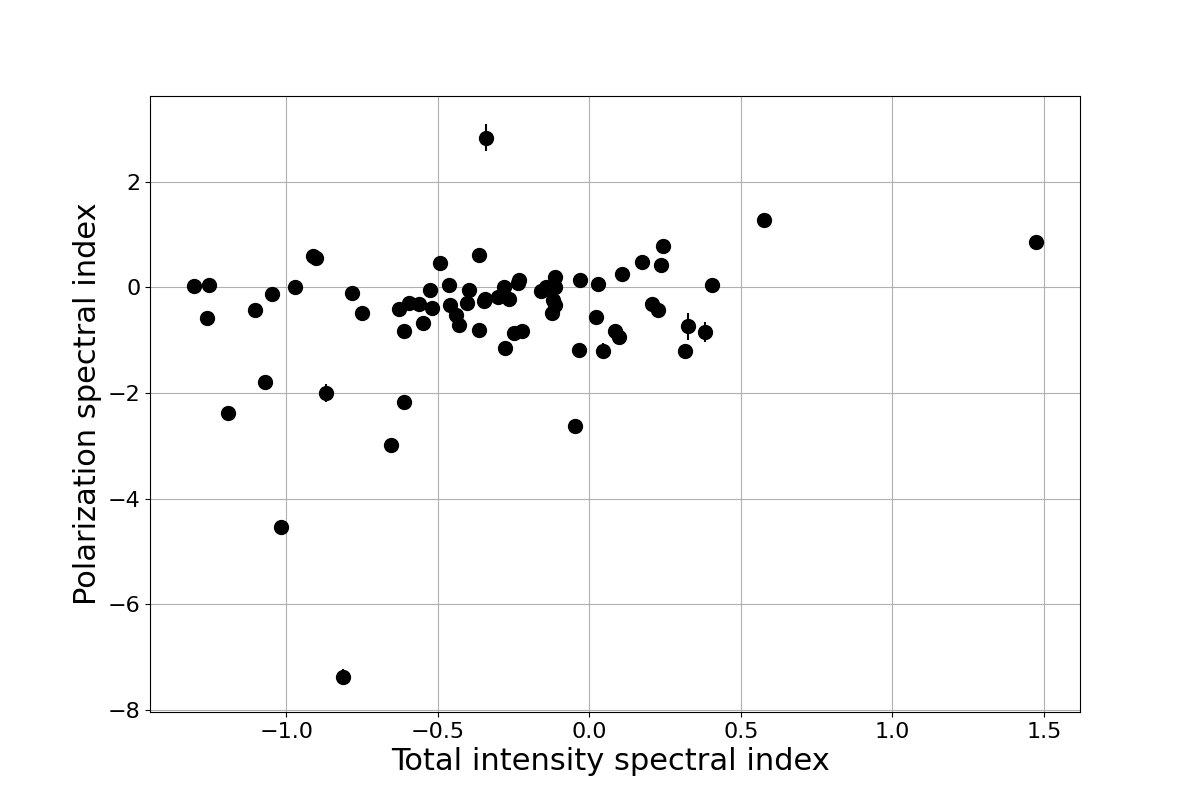}
    \caption{Fractional Polarization spectral index versus total intensity spectral index for
    69 source detected in polarization.}
    \label{fig:alphabeta}
\end{figure}

The distribution of polarized intensity $p_{\mathrm med}$, for all sources is shown in Figure~\ref{fig:pdist}.  The distribution is characterized by a broad tail of polarized sources
extending up to nearly 10\%.   The peak that is near zero corresponds to sources with
no detectable polarization.  We take 0.2\% as the minimum detectable polarization. 
We find that 69 of the calibrators sources are detected in polarization above 0.2\%.
For the polarized sources we calculated the polarization spectral index $\beta$ as 
\begin{equation}
    \label{eqn:beta}
    p \propto \lambda^{\beta}
\end{equation}
When $\beta < 0$ a source is classed as ``depolarized", i.e. the fractional polarization is reduced
at longer wavelengths (lower frequencies).  For $\beta > 0$ a source is ``repolarized'', with
fractional polarization increasing toward longer wavelengths. 
Figure~\ref{fig:alphabeta} shows a plot of the total intensity spectral index $\alpha$ versus $\beta$.
The figure shows that for flat spectrum sources ($\alpha > -0.5$) the polarized spectral index $\beta$ is typically flat.  Strong depolarization is present predominantly for steep spectrum sources.  
This result confirms a result first reported in the study of the polarized SEDs of 951 sources from published polarization measurements by \cite{Farnes_2014}.  The distinction in the
polarization properties of steep and flat spectrum sources is taken to demonstrate that 
depolarization properties of sources are due to local source environments.

\subsubsection{Potential Wide-band Polarization Calibrators}
Several of the sources show fairly strong fractional polarization.  There are 30 sources with $p_{\mathrm med} > 2.0\%$.   Many of these have a dominant Faraday
Synthesis component at RM of several 10's of rad\,m$^{-2}$.   
Calibration sources with high rotation measure
complicate polarization calibration over wide-bands at GHz frequencies and below.
For the MeerKAT L-band an RM of 10 rad\,m$^{-2}$ produces a rotation of the polarization angle across the band of 48$^{\circ}$ and rotates flux between
Stokes $Q$ and $U$ over the band.  The primary polarization calibrators, 3C\,286 and
3C\,138, have Rotation Measures very close to zero \citep{PB_2013}. \\

\begin{table}[htb]
\small
    \centering
    \begin{tabular}{lcccr}
     \hline
~~~~Name & $S_{1.4}$ & $p_{1400}$ &  pa$_{1400}$ & RM~~~~~ \\
& (Jy) & (\%) &  ($^{\circ}$) & (rad\,m$^{-2}$) \\
      \hline
    J0059+0006 & 2.45 & 3.76 &  ~67.5 & -5.0 $\pm$  0.1~~ \\
    J0108+0134 & 3.11 & 3.88 & -86.5 & -8.1 $\pm$  0.1~~ \\
    J1051-2023 & 1.44 & 2.30 & ~59.8 & -5.6 $\pm$  0.1~~ \\
    J1239-1023 & 1.55 & 2.51 & ~73.8 &  1.4  $\pm$ 0.1~~ \\
    J1424-4913 & 8.13 & 1.84 & ~-10.0 & 9.6 $\pm$ 0.1~~ \\
    J1512-0906 & 2.44 & 3.24 & ~36.8 & -11.4 $\pm$ 0.1~~ \\
    J1517-2422 & 3.03 & 3.35 & ~35.6 & -6.6  $\pm$ 0.1~~ \\
    J1550+0527 & 2.86 & 1.77 & ~73.1 & -7.4  $\pm$ 0.1~~ \\
    J1923-2104 & 1.22 & 2.93 & -38.7 & 8.5  $\pm$ 0.1~~ \\
    J2131-1207 & 1.97 & 1.75 & -49.9 & 5.5 $\pm$ 0.1~~ \\
    \hline
    \end{tabular}
    \caption{Potential Wide-band polarization calibrators.}
    \label{tab:goodcals}
\end{table}

\begin{figure*}[ht]
    \centering
    \includegraphics[width=0.9\textwidth]{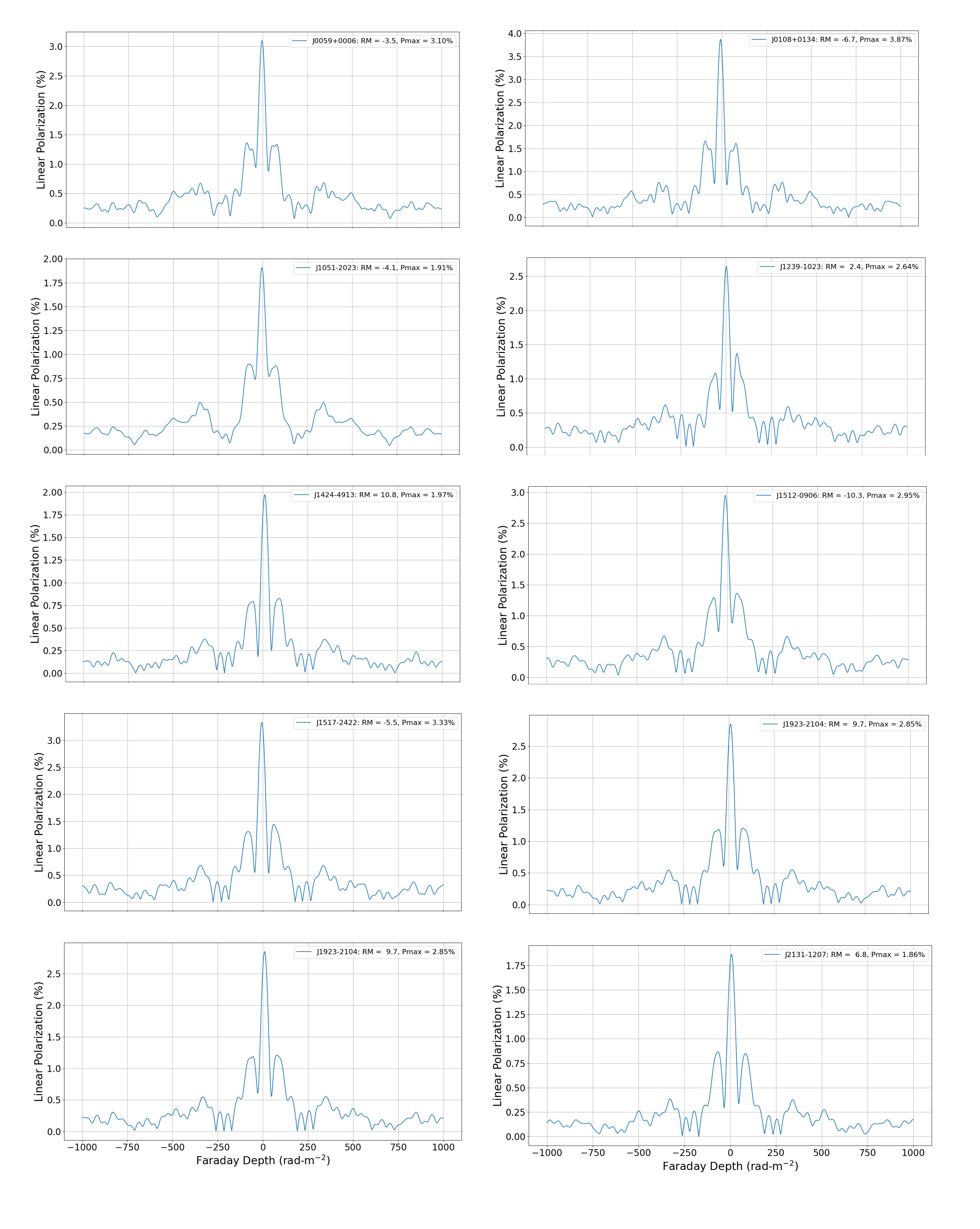}
    \caption{RM synthesis amplitude spectra for the ten sources in Table~\ref{tab:goodcals}.
    The spectra are dominated by a single RM component close to zero.
    The high sidelobes are caused by the missing frequency ranges 
    (see for example Fig.~\ref{fig:J1939_residuals}.)}
    \label{fig:RMsynth}
\end{figure*}

Table~\ref{tab:goodcals} lists ten objects with fractional polarization greater than 1.5\% that
have Faraday simple spectra and the 
Faraday depth of the dominant 
Faraday Synthesis component less than $\sim$10 rad\,m$^{-2}$. The table
lists the J2000 source name, the total intensity flux density at 1400\,MHz, 
the percent polarization and position angle at 1400\,MHz, and the
Rotation Measure of the dominant Faraday Synthesis component.
Figure~\ref{fig:RMsynth} shows the amplitude Faraday depth spectra for each,
showing a single Faraday-thin component. 
If these objects have stable polarization properties, they would be useful
calibrators for broad-band, low-frequency observations.

\subsubsection{Comparison to NVSS and SPASS}

\begin{figure*}
    \centering
    \includegraphics[width=0.9\textwidth]{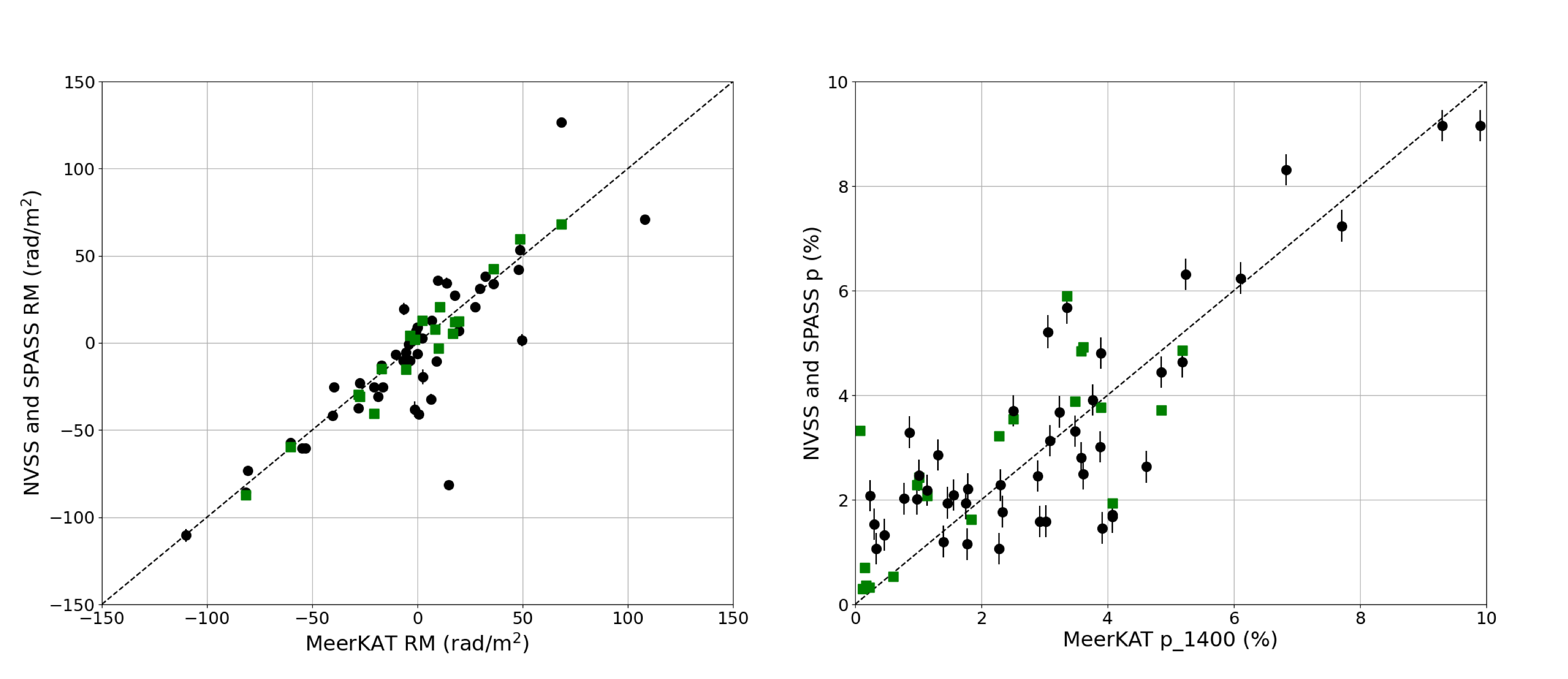}
    \caption{Left panel: Faraday depth of the dominant RM synthesis component for the MeerKAT calibrators compared to the RM from the NVSS (black circles) and dominant Faraday depth
    component from SPASS (green squares). 
    Right panel: The percent polarization from 
    MeerKAT compared to NVSS (black) and SPASS (green).}
    \label{fig:nvss_spass_compare}
\end{figure*}

Among the MeerKAT 98 calibrator sources, 54 are present in the NVSS RM catalog \citep{Taylor_2009} 
and 26 are found in the SPASS catalogue \citep{2017PASA...34...13M}. Figure~\ref{fig:nvss_spass_compare} shows a comparison of the MeerKAT results to the NVSS
and SPASS. 
The left hand panel of the figure shows the Faraday depth of the dominant component in the RM synthesis spectra of the MeerKAT calibrators compared to the Faraday Rotation Measure observed with the NVSS at 1.4\,GHz and the 
Faraday depth of the dominant component in SPASS at 2.3\,GHz. To avoid
large errors on the NVSS we include only NVSS
sources with percent polarization greater than 1\%. 

The SPASS data show very good agreement to the MeerKAT results.   
The NVSS also shows generally good agreement, but with a few outliers.
The NVSS observations \citep{1998AJ....115.1693C} were taken almost 30 years ago.  It is
not surprising to see some sources vary in RM, particularly given the compact nature of the 
calibrator source list.  We note also that the NVSS RMs were derived from a simple two-point fit to the slope of the position angle with frequency, 
which will provide a less reliable measure in the presence of Faraday complexity.

The right hand panel of Figure~\ref{fig:nvss_spass_compare} shows 
MeerKAT percent polarization at 1400\,MHz, p$_{1400}$, compared to the NVSS and the SPASS 2.3 GHz polarization. 
Error bars on the NVSS polarimetry include the estimated 0.3\% systematic error on the NVSS wide-field leakage correction \citep{Condon199NVSS}. While there is significant scatter in the data, indicating variability in polarization, there is an overall good correlation.  The SPASS polarization tends to be higher on average as expected from depolarization. 
The average SPASS to MeerKAT to depolarization for our sample is
\begin{equation}
    D = \frac{\pi_{SP}}{\pi_{MK}} = 1.35,
\end{equation}
which is similar to values found by \cite{Lamee_2016} for 416 sources detected in both the SPASS and NVSS.

\subsection{Circular Polarization}
The linear feeds of MeerKAT provide the possibility of very precise measurements
of circular polarization, as the circular polarized signals appear only in 
difference of the the cross-hand correlations (Equations 2 and 3) and is 
contaminated only by the residual polarization leakage and cross-hand phase errors.  
Circular polarization of extragalactic sources is difficult to measure due to the typically very low fractional polarization. 
As described in section~\ref{sec:leakage}, we use the strong unpolarized 
source J1939-6342 to measure the leakage and the frequency-dependence of the 
$x$ and $y$ phases independently.  The absolute cross-hand ($x-y$) phase difference is calibrated using J1331+3030. 
After applying the calibration, the residual 
band-averaged Stokes $V$ signal on J1331+3030 is 0.058\%. This value is consistent
with the residual $x-y$ phase error discussed in section~\ref{sec:cal}.
Assuming no circular polarization for J1331+3030, we can conservatively
take this as an estimate of the residual instrumental error on fractional 
circular polarization.   This value is also consistent with the maximum
residual circular polarization of J1939-6342 (Fig.~\ref{fig:J1939_residuals}).

We measure the circular polarized signal
from each source by calculating the median
amplitude of $v = V/I$ across the band, $v_{\rm med}$.  
Since circular polarization is a signed quantity, $v_{\rm med}$ may be reduced 
in the presence of real signal if $V$ changes sign across the band.  To mitigate against
this effect we also calculate the value of
$v$ at 1400 MHz, by fitting the $v$ spectra
with Equation~\ref{eqn:specfit} and solving
for $v_{\rm 1400}$.

\begin{figure}
    \centering
    \includegraphics[width=\columnwidth]{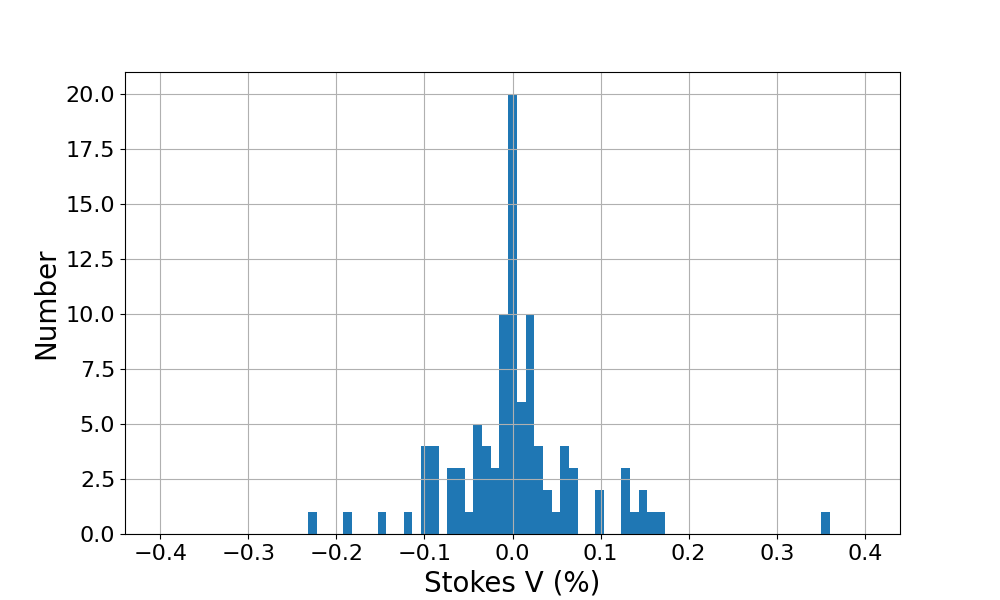}
    \caption{Distribution of band-averaged fractional Stokes $V$ (\%) for 
    the MeerKAT calibrator sources.}
    \label{fig:vdist}
\end{figure}

As a first step, for each observing run, we
compute the median $v(\nu)$ spectrum over all sources. Subsequently, if a source has $v(\nu) < 0.2\%$, we subtract the resulting median $v$ spectrum
from the source spectrum before measuring
$v_{\rm med}$ and $v_{\rm 1400}$.
Under the assumption that the median circular 
polarization of the entire ensemble of sources
will stochastically approach zero, this will reduce the effects of any residual instrumental spectrum.
 Figure~\ref{fig:vdist} shows the distribution of fractional Stokes $V$ for
the MeerKAT calibrator sources.   The distribution is centrally peaked around
zero with several outliers. One source, J0240-2309 is very strongly circularly
polarized with $V$ close to 0.4\%.  The standard deviation of the distribution, 
measured as 1.486 times the median absolute deviation (MAD), is 0.062\%, and the width 
of the central peak is $\sim$0.2\%.
We conservatively take 0.07\% as the detection
threshold for circular polarization.

 \begin{figure}
    \centering
    \includegraphics[width=\columnwidth]{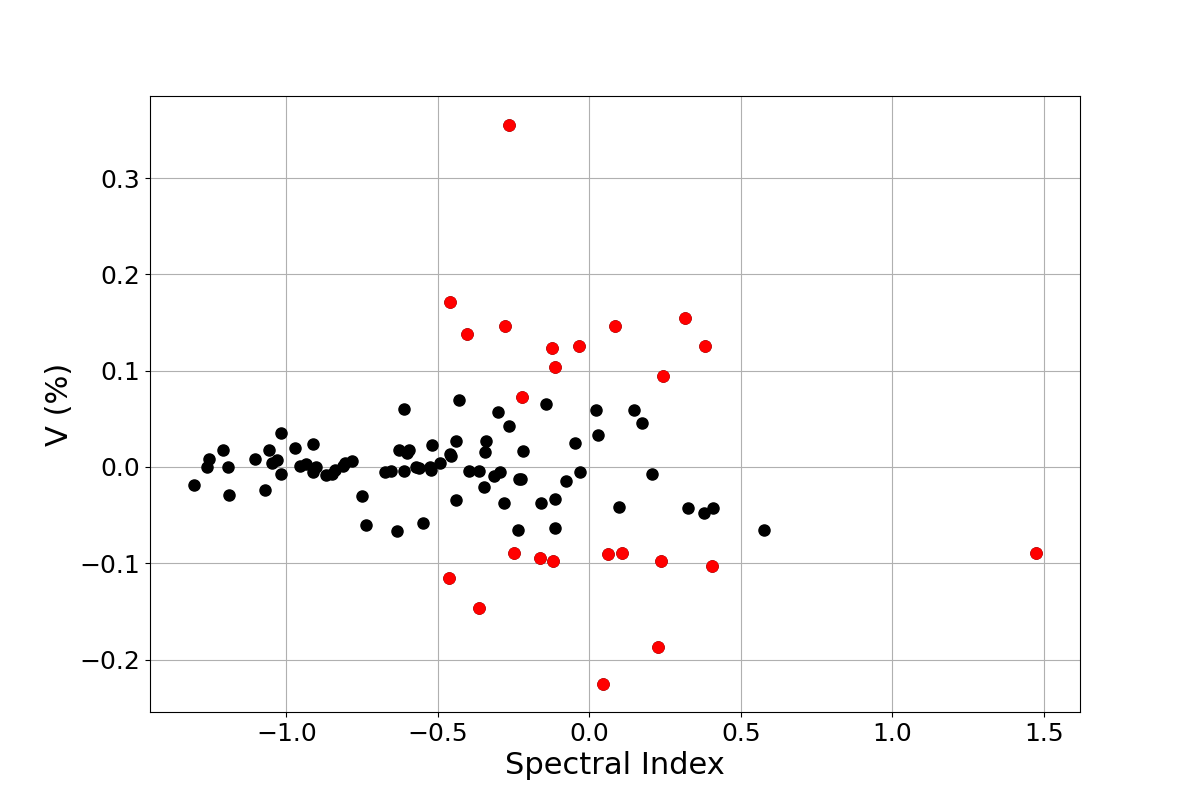}
    \caption{Fractional band-averaged circular polarization (\%) versus
    the MeerKAT spectral index from the Stokes $I$ spectrum. Sources with
    measured values of $v_{\rm med}$$ > 0.07$\% are shown in red.}
    \label{fig:alpha_v}
\end{figure}

Cicular polarization is detected with 
$v_{\rm med}$ above 0.07\% from 24 of the calibrators. 
Figure~\ref{fig:alpha_v} plots $v_{\rm med}$ versus the total intensity spectral index, $\alpha$, determined by the fit to the
Stokes $I$ spectrum.  Source with $V > 0.07\%$ are shown in red. 
Circular polarization is detected in 24\% of the sources.
It is striking that significant circular polarization is detected 
only for flat spectrum sources with 
$\alpha > -0.5$. Steep spectrum sources in the sample do not exhibit circular polarization.
This is consistent with the result of high precision circular polarization measurements 
of 31 radio sources at 5\,GHz with the ATCA \citep{Rayner_2000}, 
which also showed circular polarization only for sources
with spectral index $\alpha > -0.5$ between wavelengths of 3\,cm$-$20\,cm.

\section{Data Availability}

The $IQUV$ image cubes and spectro-polarimetric data for each source are available on the science archive of the
South African Radio Astronomy Observatory under DOI \url{https://doi.org/10.48479/sbtr-k883}.
The site also contains plots of frequency spectra of total intensity, the Stokes $Q$ and $U$ versus $\lambda^2$, the Faraday synthesis spectra, and linear and circular polarization spectra for each source.

A summary of the polarization properties of the calibration sources is given in Table~\ref{tab:sourcetable} in Appendix A.
The columns of the table lists the following properties for each source.
\begin{enumerate}
  \item The J2000 coordinate source name. 
  \item The flux density at 1400\,MHz ($S_{1.4}$) with error, from a power-law fit to the Stokes $I$ spectrum.
  \item The in-band spectral index and error from the spectral fit (Equation~\ref{eqn:specfit}) to the Stokes $I$ spectrum ($\alpha$). Errors for both the flux density and spectral index are the formal error on the fit parameters from the covariance matrix.
  \item The curvature term and error from the spectral fit to the Stokes $I$ spectrum ($\alpha$). Errors are from the covariance matrix.
  \item The percent polarized intensity and position angle at 1400\,MHz (p$_{1400}$, pa$_{1400}$) measured as the average value of polarized intensity and polarization angle over the channels within $\pm$20\,MHz of 1400\,MHz.  
  \item The spectral index of polarized intensity ($\beta$) from a power-law fit to the spectrum of polarized intensity (Equation~\ref{eqn:beta}). Here also the errors are the formal error on the fit parameter from the covariance matrix.
  \item The Faraday depth of the dominant RM synthesis component with error.
  \item The amplitude (percent polarization) of the dominant RM synthesis component.
  \item The median value fractional Stokes $V$ over the band, $v_{\rm med}$.  The error is the  standard error on the mean derived from the RMS of the $N$ values over the frequency interval divided by $\sqrt{N}$.
  \item The circular polarization at 1400\,MHz, $v_{1400}$, with error, from the spectral fit to the $v$ spectrum.
\end{enumerate}
The errors listed in Appendix A include only the random error from the noise of the data.  The spectral variance is much smaller than the residual instrumental leakage error, which is estimated as 0.16\% in $p$ and 0.05\% in $v$.   
Detection thresholds have been taken as 0.2\% for $p$ and 0.07\% for $v$.   The systematic error on Rotation Measure is 
0.3 rad\,m$^{-2}$ and on position angle is 2$^{\circ}$.
The systematic error in flux density is estimated to be 3\%, taken as the fractional difference in the derived flux of 3C 286 between the two runs.  These systematic uncertainties should be added in quadrature with the errors in the Appendix to derive the total error.

The MeerKAT program of polarization observations
of potential calibrators is ongoing. A monitoring program of the ten sources in Table~\ref{tab:goodcals} is in progress to assess their potential for this purpose. This MeerKAT program observes the ten sources every $\sim$3 months and has been collecting both L-band and UHF (580 - 1015 MHz corresponding to 544 - 1088 MHz digitized) data since March 2021.   These results will be published in a following publication. Similarly, there is an ongoing campaign to characterize the MeerKAT UHF and S-band calibrator sources in full polarization. The S-band campaign is in the early stages of identifying additional suitable MeerKAT S-band calibrator sources whose distribution on the sky covers a larger area than the existing list of S-band calibrators. The UHF campaign, on the other hand, is in the advanced data collection stage with reductions to begin in early 2024.\\

\begin{acknowledgments}
{\it Acknowledgments}. The MeerKAT telescope is operated by the South African Radio Astronomy Observatory, which is a facility of the National Research Foundation, an agency of the Department of Science and Innovation.
This work was carried out using the data processing pipelines developed at the Inter-University Institute for Data Intensive Astronomy (IDIA) and available at \url{https://idia-pipelines.github.io}. IDIA is a partnership of the University of Cape Town, the University of Pretoria, the University of the Western Cape.
We acknowledge the use of the ilifu cloud computing facility – \url{www.ilifu.ac.za}, a partnership between the University of Cape Town, the University of the Western Cape, the University of Stellenbosch, Sol Plaatje University, the Cape Peninsula University of Technology and the South African Radio Astronomy Observatory. The Ilifu facility is supported by contributions from the Inter-University Institute for Data Intensive Astronomy, the Computational Biology division at UCT and the Data Intensive Research Initiative of South Africa (DIRISA).
\end{acknowledgments}

\bibliography{main}{}
\bibliographystyle{aasjournal}

%




\appendix

\section{Summary Table of Polarization properties}

\startlongtable
\begin{longrotatetable}
\begin{deluxetable}{lrrrrrrrrrr}
\tabletypesize{\scriptsize}
\tablecaption{Polarization Properties of MeerKAT Calibrator Sources.}
\label{tab:sourcetable} 
\tablehead{
\colhead{Name} & 
\colhead{S$_{1.4}$} & 
\colhead{$\alpha$} &
\colhead{$C$} &
\colhead{p$_{1400}$} &
\colhead{pa$_{1400}$} &
\colhead{$\beta$} &
\colhead{RM} &
\colhead{p$_{\mathrm{max}}$} &
\colhead{v$_{med}$}  &
\colhead{v$_{1400}$}\\
\colhead{} & 
\colhead{$(\mathrm{Jy})$} & 
\colhead{} &
\colhead{} &
\colhead{(\%)} &
\colhead{$(\mathrm{deg})$} &
\colhead{} &
\colhead{(rad\,m$^{-2}$)} &
\colhead{(\%)} &
\colhead{(\%)}  &
\colhead{(\%)}
}
\startdata





J0010-4153 &    4.562 $\pm$  0.001 &    -0.914 $\pm$ 0.002 &    -0.226 $\pm$ 0.007 &    0.12 $\pm$  0.002 &   -67.4 $\pm$  0.8 &    -0.24 $\pm$  0.18 &    -2.8 $\pm$   0.2 &   0.14~~ &    0.024 $\pm$  0.002 &    0.020 $\pm$  0.002\\
J0022+0014 &    2.913 $\pm$  0.001 &    -0.634 $\pm$ 0.003 &    -0.250 $\pm$ 0.008 &    0.09 $\pm$  0.002 &   -18.3 $\pm$  1.7 &    -1.53 $\pm$  0.23 &     0.5 $\pm$   0.2 &   0.13~~ &   -0.066 $\pm$  0.001 &   -0.067 $\pm$  0.002\\
J0024-4202 &    2.914 $\pm$  0.001 &     0.407 $\pm$ 0.003 &    -0.830 $\pm$ 0.012 &    0.09 $\pm$  0.002 &   -11.3 $\pm$  1.1 &    -0.36 $\pm$  0.20 &     3.0 $\pm$   0.2 &   0.13~~ &   -0.043 $\pm$  0.002 &   -0.056 $\pm$  0.002\\
J0025-2602 &    8.734 $\pm$  0.002 &    -0.738 $\pm$ 0.002 &    -0.017 $\pm$ 0.006 &    0.17 $\pm$  0.001 &    65.7 $\pm$  1.2 &     0.31 $\pm$  0.13 &    -5.2 $\pm$   0.1 &   0.18~~ &   -0.060 $\pm$  0.003 &   -0.062 $\pm$  0.004\\
J0059+0006 &    2.447 $\pm$  0.001 &    -0.548 $\pm$ 0.003 &     0.041 $\pm$ 0.008 &    3.76 $\pm$  0.003 &    67.5 $\pm$  0.6 &    -0.68 $\pm$  0.03 &    -5.0 $\pm$   0.1 &   3.10~~ &   -0.053 $\pm$  0.001 &   -0.046 $\pm$  0.006\\
J0108+0134 &    3.106 $\pm$  0.001 &    -0.232 $\pm$ 0.004 &     0.144 $\pm$ 0.011 &    3.88 $\pm$  0.003 &   -86.5 $\pm$  0.2 &     0.13 $\pm$  0.01 &    -8.1 $\pm$   0.1 &   3.87~~ &   -0.007 $\pm$  0.001 &    0.112 $\pm$  0.020\\
J0137+3309 &   16.096 $\pm$  0.004 &    -0.869 $\pm$ 0.002 &     0.067 $\pm$ 0.005 &    0.64 $\pm$  0.003 &   -43.2 $\pm$  0.6 &    -2.01 $\pm$  0.17 &   -56.8 $\pm$   0.1 &   0.49~~ &   -0.009 $\pm$  0.001 &   -0.004 $\pm$  0.002\\
J0155-4048 &    2.161 $\pm$  0.001 &    -0.458 $\pm$ 0.003 &    -0.010 $\pm$ 0.008 &    0.07 $\pm$  0.003 &    22.7 $\pm$  2.3 &     1.72 $\pm$  0.13 &    -1.6 $\pm$   0.3 &   0.08~~ &    0.011 $\pm$  0.001 &    0.012 $\pm$  0.002\\
J0203-4349 &    2.725 $\pm$  0.001 &    -0.815 $\pm$ 0.003 &     0.036 $\pm$ 0.008 &    0.91 $\pm$  0.002 &   -62.7 $\pm$  0.3 &    -7.37 $\pm$  0.14 &    -9.8 $\pm$   0.1 &   0.49~~ &    0.001 $\pm$  0.001 &   -0.000 $\pm$  0.001\\
J0210-5101 &    3.385 $\pm$  0.001 &     0.206 $\pm$ 0.003 &     0.276 $\pm$ 0.011 &    1.24 $\pm$  0.002 &    23.4 $\pm$  0.2 &    -0.31 $\pm$  0.02 &    13.0 $\pm$   0.1 &   1.12~~ &   -0.005 $\pm$  0.001 &   -0.053 $\pm$  0.005\\
J0238+1636 &    0.524 $\pm$  0.001 &    -0.114 $\pm$ 0.012 &     0.466 $\pm$ 0.035 &    0.77 $\pm$  0.013 &    81.9 $\pm$  0.9 &     0.20 $\pm$  0.10 &    46.6 $\pm$   0.1 &   0.72~~ &   -0.062 $\pm$  0.003 &   -0.113 $\pm$  0.007\\
J0240-2309 &    5.978 $\pm$  0.002 &    -0.266 $\pm$ 0.003 &    -0.401 $\pm$ 0.008 &    0.98 $\pm$  0.002 &   -49.2 $\pm$  0.1 &    -0.21 $\pm$  0.05 &     8.5 $\pm$   0.1 &   0.98~~ &    0.353 $\pm$  0.001 &    0.307 $\pm$  0.002\\
J0252-7104 &    5.810 $\pm$  0.001 &    -0.971 $\pm$ 0.002 &     0.004 $\pm$ 0.005 &    0.18 $\pm$  0.001 &   -42.8 $\pm$  0.2 &     0.01 $\pm$  0.10 &     0.7 $\pm$   0.2 &   0.23~~ &    0.018 $\pm$  0.001 &    0.019 $\pm$  0.001\\
J0303-6211 &    3.198 $\pm$  0.001 &    -0.112 $\pm$ 0.004 &    -0.066 $\pm$ 0.011 &    4.71 $\pm$  0.002 &    39.2 $\pm$  0.5 &    -0.34 $\pm$  0.01 &    47.9 $\pm$   0.1 &   4.43~~ &   -0.037 $\pm$  0.001 &   -0.032 $\pm$  0.006\\
J0318+1628 &    7.653 $\pm$  0.002 &    -0.602 $\pm$ 0.002 &    -0.260 $\pm$ 0.007 &    0.03 $\pm$  0.002 &   -15.8 $\pm$ 10.4 &    -2.22 $\pm$  0.34 &     3.5 $\pm$   0.4 &   0.05~~ &    0.015 $\pm$  0.001 &    0.001 $\pm$  0.001\\
J0323+0534 &    2.764 $\pm$  0.001 &    -0.911 $\pm$ 0.004 &     0.029 $\pm$ 0.010 &    0.26 $\pm$  0.003 &   -78.6 $\pm$  0.3 &     0.59 $\pm$  0.12 &     0.9 $\pm$   0.1 &   0.33~~ &   -0.006 $\pm$  0.001 &   -0.007 $\pm$  0.002\\
J0329+2756 &    1.397 $\pm$  0.001 &    -0.365 $\pm$ 0.004 &    -0.104 $\pm$ 0.012 &    0.16 $\pm$  0.005 &   -10.7 $\pm$  1.2 &    -0.82 $\pm$  0.14 &     0.9 $\pm$   0.2 &   0.21~~ &   -0.148 $\pm$  0.001 &   -0.133 $\pm$  0.002\\
J0403+2600 &    1.295 $\pm$  0.001 &    -0.033 $\pm$ 0.004 &    -0.248 $\pm$ 0.012 &    3.05 $\pm$  0.005 &    68.6 $\pm$  0.6 &    -1.19 $\pm$  0.01 &    48.4 $\pm$   0.1 &   2.62~~ &    0.131 $\pm$  0.002 &    0.111 $\pm$  0.008\\
J0405-1308 &    3.936 $\pm$  0.002 &    -0.494 $\pm$ 0.004 &     0.140 $\pm$ 0.011 &    1.14 $\pm$  0.002 &    24.4 $\pm$  0.3 &     0.46 $\pm$  0.04 &    18.1 $\pm$   0.1 &   1.18~~ &    0.006 $\pm$  0.001 &    0.004 $\pm$  0.002\\
J0408-6545 &   15.207 $\pm$  0.002 &    -1.210 $\pm$ 0.001 &    -0.033 $\pm$ 0.003 &    0.01 $\pm$  0.000 &   -72.7 $\pm$  0.8 &     1.05 $\pm$  0.10 &    -0.5 $\pm$   0.3 &   0.02~~ &    0.018 $\pm$  0.001 &    0.022 $\pm$  0.001\\
J0409-1757 &    2.203 $\pm$  0.001 &    -0.841 $\pm$ 0.004 &     0.047 $\pm$ 0.012 &    0.15 $\pm$  0.003 &   -42.3 $\pm$  1.0 &     0.63 $\pm$  0.10 &     3.6 $\pm$   0.1 &   0.20~~ &   -0.004 $\pm$  0.001 &   -0.007 $\pm$  0.002\\
J0420-6223 &    3.323 $\pm$  0.001 &    -1.030 $\pm$ 0.003 &     0.069 $\pm$ 0.009 &    0.12 $\pm$  0.002 &    58.3 $\pm$  0.4 &     0.00 $\pm$  0.10 &     0.9 $\pm$   0.2 &   0.16~~ &    0.007 $\pm$  0.001 &    0.009 $\pm$  0.002\\
J0423-0120 &    1.187 $\pm$  0.001 &     0.029 $\pm$ 0.005 &     0.417 $\pm$ 0.014 &    4.07 $\pm$  0.005 &    31.9 $\pm$  0.3 &     0.06 $\pm$  0.01 &   -22.1 $\pm$   0.1 &   3.98~~ &    0.031 $\pm$  0.002 &    0.026 $\pm$  0.008\\
J0440-4333 &    3.591 $\pm$  0.002 &    -0.751 $\pm$ 0.004 &     0.043 $\pm$ 0.010 &    0.60 $\pm$  0.002 &   -83.9 $\pm$  0.2 &    -0.48 $\pm$  0.02 &     6.8 $\pm$   0.1 &   0.57~~ &   -0.031 $\pm$  0.001 &   -0.019 $\pm$  0.002\\
J0447-2203 &    2.028 $\pm$  0.001 &    -1.047 $\pm$ 0.003 &    -0.065 $\pm$ 0.009 &    0.23 $\pm$  0.003 &   -31.8 $\pm$  0.5 &    -0.13 $\pm$  0.10 &     0.3 $\pm$   0.1 &   0.28~~ &    0.003 $\pm$  0.001 &    0.000 $\pm$  0.001\\
J0453-2807 &    2.192 $\pm$  0.001 &    -0.340 $\pm$ 0.004 &     0.007 $\pm$ 0.012 &    0.59 $\pm$  0.003 &     9.8 $\pm$  1.6 &     2.83 $\pm$  0.26 &    18.0 $\pm$   0.1 &   1.73~~ &    0.026 $\pm$  0.001 &    0.039 $\pm$  0.006\\
J0503+0203 &    2.233 $\pm$  0.001 &     0.148 $\pm$ 0.004 &    -0.356 $\pm$ 0.014 &    0.04 $\pm$  0.003 &    34.5 $\pm$  1.8 &     1.51 $\pm$  0.14 &     0.0 $\pm$   0.5 &   0.06~~ &    0.059 $\pm$  0.001 &    0.072 $\pm$  0.002\\
J0521+1638 &    8.320 $\pm$  0.002 &    -0.613 $\pm$ 0.002 &     0.083 $\pm$ 0.005 &    7.71 $\pm$  0.002 &   -15.1 $\pm$  0.2 &    -0.83 $\pm$  0.01 &    -2.4 $\pm$   0.1 &   7.03~~ &    0.049 $\pm$  0.001 &    0.023 $\pm$  0.007\\
J0534+1927 &    6.739 $\pm$  0.002 &    -0.849 $\pm$ 0.002 &     0.022 $\pm$ 0.006 &    0.02 $\pm$  0.001 &   -10.4 $\pm$ 11.1 &    -2.06 $\pm$  0.29 &     5.2 $\pm$   0.3 &   0.05~~ &   -0.007 $\pm$  0.001 &   -0.006 $\pm$  0.001\\
J0538-4405 &    2.142 $\pm$  0.001 &    -0.226 $\pm$ 0.004 &     0.384 $\pm$ 0.011 &    0.07 $\pm$  0.002 &    63.5 $\pm$  1.1 &     3.25 $\pm$  0.14 &    47.2 $\pm$   0.2 &   0.17~~ &   -0.013 $\pm$  0.001 &   -0.004 $\pm$  0.002\\
J0609-1542 &    1.663 $\pm$  0.001 &    -0.462 $\pm$ 0.006 &     0.356 $\pm$ 0.016 &    2.28 $\pm$  0.003 &     2.1 $\pm$  0.9 &     0.05 $\pm$  0.03 &    66.8 $\pm$   0.1 &   2.45~~ &   -0.117 $\pm$  0.001 &   -0.116 $\pm$  0.003\\
J0616-3456 &    2.902 $\pm$  0.001 &    -0.571 $\pm$ 0.003 &     0.042 $\pm$ 0.008 &    0.13 $\pm$  0.002 &    84.8 $\pm$  0.4 &     0.22 $\pm$  0.09 &    -0.2 $\pm$   0.2 &   0.16~~ &   -0.001 $\pm$  0.001 &   -0.018 $\pm$  0.003\\
J0632+1022 &    2.441 $\pm$  0.001 &    -0.526 $\pm$ 0.003 &    -0.326 $\pm$ 0.008 &    0.24 $\pm$  0.002 &    30.6 $\pm$  0.9 &    -0.06 $\pm$  0.10 &     1.3 $\pm$   0.1 &   0.27~~ &    0.001 $\pm$  0.001 &    0.032 $\pm$  0.003\\
J0725-0054 &   11.809 $\pm$  0.004 &     0.228 $\pm$ 0.003 &    -0.374 $\pm$ 0.009 &    2.52 $\pm$  0.003 &   -19.5 $\pm$  0.6 &    -0.43 $\pm$  0.01 &    47.6 $\pm$   0.1 &   2.39~~ &   -0.187 $\pm$  0.001 &   -0.204 $\pm$  0.003\\
J0730-1141 &    2.246 $\pm$  0.002 &     0.576 $\pm$ 0.008 &     0.366 $\pm$ 0.026 &    0.98 $\pm$  0.003 &   -79.6 $\pm$  1.7 &     1.27 $\pm$  0.10 &   106.7 $\pm$   0.1 &   1.49~~ &   -0.069 $\pm$  0.001 &   -0.064 $\pm$  0.002\\
J0735-1735 &    2.598 $\pm$  0.001 &    -0.162 $\pm$ 0.004 &     0.039 $\pm$ 0.013 &    0.06 $\pm$  0.002 &    43.1 $\pm$  2.3 &    -1.40 $\pm$  0.20 &    -1.5 $\pm$   0.4 &   0.07~~ &   -0.095 $\pm$  0.001 &   -0.088 $\pm$  0.001\\
J0739+0137 &    1.014 $\pm$  0.001 &     0.243 $\pm$ 0.004 &     0.386 $\pm$ 0.013 &    5.24 $\pm$  0.006 &   -52.2 $\pm$ 18.9 &     0.79 $\pm$  0.01 &    26.3 $\pm$   0.1 &   5.81~~ &    0.092 $\pm$  0.001 &    0.072 $\pm$  0.005\\
J0745+1011 &    3.254 $\pm$  0.002 &     0.382 $\pm$ 0.004 &    -0.484 $\pm$ 0.015 &    0.22 $\pm$  0.004 &    40.3 $\pm$  0.4 &    -0.85 $\pm$  0.19 &     2.0 $\pm$   0.2 &   0.27~~ &    0.126 $\pm$  0.001 &    0.129 $\pm$  0.001\\
J0825-5010 &    6.309 $\pm$  0.002 &     0.377 $\pm$ 0.002 &    -0.576 $\pm$ 0.008 &    0.23 $\pm$  0.001 &    53.9 $\pm$  2.2 &     1.30 $\pm$  0.19 &    -0.2 $\pm$   0.1 &   0.12~~ &   -0.047 $\pm$  0.001 &   -0.071 $\pm$  0.001\\
J0828-3731 &    2.090 $\pm$  0.002 &    -0.282 $\pm$ 0.007 &    -0.075 $\pm$ 0.022 &    0.35 $\pm$  0.003 &    63.3 $\pm$ 17.3 &     0.01 $\pm$  0.09 &     0.0 $\pm$   0.1 &   0.45~~ &   -0.037 $\pm$  0.001 &   -0.025 $\pm$  0.002\\
J0842+1835 &    1.037 $\pm$  0.001 &    -0.278 $\pm$ 0.007 &     0.098 $\pm$ 0.020 &    3.09 $\pm$  0.006 &     7.9 $\pm$  0.3 &    -1.15 $\pm$  0.03 &    31.0 $\pm$   0.1 &   2.60~~ &    0.147 $\pm$  0.001 &    0.119 $\pm$  0.003\\
J0854+2006 &    2.048 $\pm$  0.001 &     0.084 $\pm$ 0.003 &    -0.021 $\pm$ 0.009 &    6.83 $\pm$  0.005 &    24.0 $\pm$  0.3 &    -0.82 $\pm$  0.01 &    28.8 $\pm$   0.1 &   6.05~~ &    0.150 $\pm$  0.001 &    0.179 $\pm$  0.009\\
J0906-6829 &    1.817 $\pm$  0.001 &    -1.103 $\pm$ 0.003 &     0.012 $\pm$ 0.007 &    2.27 $\pm$  0.003 &   -34.6 $\pm$  0.5 &    -0.44 $\pm$  0.03 &   -49.8 $\pm$   0.1 &   2.06~~ &    0.008 $\pm$  0.001 &    0.008 $\pm$  0.002\\
J1008+0730 &    6.526 $\pm$  0.001 &    -0.902 $\pm$ 0.002 &     0.061 $\pm$ 0.005 &    0.22 $\pm$  0.002 &   -11.9 $\pm$  0.2 &     0.56 $\pm$  0.07 &    -0.8 $\pm$   0.1 &   0.30~~ &   -0.000 $\pm$  0.001 &   -0.000 $\pm$  0.001\\
J1051-2023 &    1.442 $\pm$  0.001 &    -1.072 $\pm$ 0.003 &     0.009 $\pm$ 0.009 &    2.30 $\pm$  0.004 &    59.8 $\pm$  0.1 &    -1.80 $\pm$  0.03 &    -5.6 $\pm$   0.1 &   1.91~~ &   -0.021 $\pm$  0.001 &   -0.024 $\pm$  0.003\\
J1058+0133 &    3.665 $\pm$  0.001 &    -0.222 $\pm$ 0.003 &     0.118 $\pm$ 0.008 &    4.07 $\pm$  0.003 &   -17.9 $\pm$  0.5 &    -0.83 $\pm$  0.02 &   -41.0 $\pm$   0.1 &   3.65~~ &    0.074 $\pm$  0.001 &    0.082 $\pm$  0.007\\
J1120-2508 &    1.640 $\pm$  0.001 &    -0.613 $\pm$ 0.007 &    -0.053 $\pm$ 0.023 &    1.46 $\pm$  0.004 &   -37.2 $\pm$  0.2 &    -2.18 $\pm$  0.05 &     8.0 $\pm$   0.1 &   1.14~~ &   -0.005 $\pm$  0.001 &   -0.006 $\pm$  0.002\\
J1130-1449 &    4.842 $\pm$  0.003 &    -0.250 $\pm$ 0.006 &    -0.061 $\pm$ 0.018 &    4.85 $\pm$  0.003 &    66.3 $\pm$  0.5 &    -0.87 $\pm$  0.02 &    35.2 $\pm$   0.1 &   4.27~~ &   -0.093 $\pm$  0.001 &   -0.095 $\pm$  0.005\\
J1154-3505 &    6.088 $\pm$  0.004 &    -0.523 $\pm$ 0.006 &    -0.048 $\pm$ 0.019 &    0.16 $\pm$  0.004 &    43.8 $\pm$  1.2 &     1.70 $\pm$  0.16 &    -5.9 $\pm$   0.2 &   0.19~~ &   -0.004 $\pm$  0.001 &   -0.003 $\pm$  0.002\\
J1215-1731 &    1.695 $\pm$  0.001 &    -0.345 $\pm$ 0.007 &     0.195 $\pm$ 0.021 &    2.01 $\pm$  0.004 &     6.9 $\pm$  0.2 &    -0.22 $\pm$  0.02 &   -15.7 $\pm$   0.1 &   1.88~~ &    0.015 $\pm$  0.001 &    0.024 $\pm$  0.003\\
J1239-1023 &    1.558 $\pm$  0.001 &    -0.112 $\pm$ 0.007 &    -0.163 $\pm$ 0.023 &    2.51 $\pm$  0.004 &    73.8 $\pm$  0.2 &    -0.00 $\pm$  0.05 &     1.4 $\pm$   0.1 &   2.64~~ &    0.106 $\pm$  0.001 &    0.099 $\pm$  0.003\\
J1246-2547 &    0.845 $\pm$  0.001 &     0.173 $\pm$ 0.007 &     0.678 $\pm$ 0.025 &    3.48 $\pm$  0.007 &   -30.6 $\pm$  0.4 &     0.48 $\pm$  0.02 &   -29.2 $\pm$   0.1 &   3.67~~ &    0.047 $\pm$  0.002 &    0.051 $\pm$  0.005\\
J1256-0547 &    9.734 $\pm$  0.013 &    -0.403 $\pm$ 0.012 &     0.329 $\pm$ 0.036 &    3.61 $\pm$  0.004 &   -44.7 $\pm$  0.2 &    -0.30 $\pm$  0.04 &    16.9 $\pm$   0.1 &   3.43~~ &    0.136 $\pm$  0.001 &    0.156 $\pm$  0.003\\
J1311-2216 &    4.834 $\pm$  0.005 &    -1.190 $\pm$ 0.010 &     0.338 $\pm$ 0.028 &    0.05 $\pm$  0.004 &    13.3 $\pm$  7.4 &    -0.64 $\pm$  0.33 &   -13.8 $\pm$   1.1 &   0.04~~ &   -0.029 $\pm$  0.001 &   -0.027 $\pm$  0.002\\
J1318-4620 &    2.205 $\pm$  0.002 &    -0.935 $\pm$ 0.008 &    -0.001 $\pm$ 0.022 &    0.04 $\pm$  0.004 &   -11.5 $\pm$ 16.1 &     0.17 $\pm$  0.25 &     0.5 $\pm$   0.5 &   0.10~~ &    0.003 $\pm$  0.001 &    0.001 $\pm$  0.001\\
J1323-4452 &    3.026 $\pm$  0.003 &    -0.808 $\pm$ 0.008 &    -0.004 $\pm$ 0.025 &    0.11 $\pm$  0.004 &    -2.6 $\pm$  2.2 &    -0.29 $\pm$  0.30 &    -0.3 $\pm$   1.3 &   0.04~~ &    0.005 $\pm$  0.001 &    0.000 $\pm$  0.001\\
J1331+3030 &   14.673 $\pm$  0.003 &    -0.630 $\pm$ 0.002 &     0.001 $\pm$ 0.005 &    9.90 $\pm$  0.003 &    29.2 $\pm$  0.1 &    -0.42 $\pm$  0.01 &    -0.3 $\pm$   0.1 &   9.39~~ &    0.016 $\pm$  0.001 &    0.029 $\pm$  0.012\\
J1337-1257 &    2.509 $\pm$  0.002 &     0.315 $\pm$ 0.007 &     0.404 $\pm$ 0.023 &    1.01 $\pm$  0.004 &    -1.6 $\pm$  0.3 &    -1.20 $\pm$  0.05 &   -18.3 $\pm$   0.1 &   0.81~~ &    0.154 $\pm$  0.001 &    0.137 $\pm$  0.001\\
J1347+1217 &    5.203 $\pm$  0.004 &    -0.459 $\pm$ 0.006 &     0.113 $\pm$ 0.018 &    0.04 $\pm$  0.004 &    55.2 $\pm$  3.5 &    -0.78 $\pm$  0.24 &     7.0 $\pm$   0.9 &   0.04~~ &    0.014 $\pm$  0.001 &    0.019 $\pm$  0.001\\
J1424-4913 &    8.123 $\pm$  0.004 &    -0.366 $\pm$ 0.005 &     0.059 $\pm$ 0.015 &    1.84 $\pm$  0.003 &   -10.0 $\pm$  0.2 &     0.61 $\pm$  0.01 &     9.6 $\pm$   0.1 &   1.97~~ &   -0.004 $\pm$  0.001 &   -0.001 $\pm$  0.001\\
J1427-4206 &    4.465 $\pm$  0.004 &    -0.049 $\pm$ 0.007 &    -0.014 $\pm$ 0.024 &    2.00 $\pm$  0.003 &   -51.6 $\pm$  0.4 &    -2.62 $\pm$  0.06 &   -40.8 $\pm$   0.1 &   1.54~~ &    0.025 $\pm$  0.001 &    0.031 $\pm$  0.002\\
J1445+0958 &    2.177 $\pm$  0.002 &    -0.461 $\pm$ 0.007 &    -0.307 $\pm$ 0.021 &    0.86 $\pm$  0.004 &   -80.7 $\pm$  0.3 &    -0.33 $\pm$  0.04 &    13.1 $\pm$   0.1 &   0.78~~ &    0.172 $\pm$  0.001 &    0.182 $\pm$  0.002\\
J1501-3918 &    2.805 $\pm$  0.003 &    -0.675 $\pm$ 0.008 &     0.011 $\pm$ 0.024 &    0.19 $\pm$  0.003 &    12.9 $\pm$  0.7 &     0.21 $\pm$  0.09 &    -0.4 $\pm$   0.2 &   0.21~~ &   -0.005 $\pm$  0.001 &   -0.004 $\pm$  0.001\\
J1512-0906 &    2.434 $\pm$  0.002 &     0.021 $\pm$ 0.006 &     0.205 $\pm$ 0.020 &    3.24 $\pm$  0.003 &    36.8 $\pm$  0.1 &    -0.57 $\pm$  0.01 &   -11.4 $\pm$   0.1 &   2.95~~ &    0.059 $\pm$  0.001 &    0.067 $\pm$  0.001\\
J1517-2422 &    3.030 $\pm$  0.002 &    -0.031 $\pm$ 0.006 &    -0.084 $\pm$ 0.019 &    3.35 $\pm$  0.003 &    35.6 $\pm$  0.2 &     0.13 $\pm$  0.03 &    -6.6 $\pm$   0.1 &   3.33~~ &   -0.005 $\pm$  0.001 &   -0.005 $\pm$  0.003\\
J1550+0527 &    2.864 $\pm$  0.002 &    -0.160 $\pm$ 0.006 &     0.008 $\pm$ 0.020 &    1.77 $\pm$  0.004 &    73.1 $\pm$  0.2 &    -0.07 $\pm$  0.01 &    -7.4 $\pm$   0.1 &   1.68~~ &   -0.036 $\pm$  0.001 &   -0.031 $\pm$  0.002\\
J1605-1734 &    1.375 $\pm$  0.001 &    -1.193 $\pm$ 0.009 &     0.028 $\pm$ 0.026 &    2.33 $\pm$  0.004 &   -54.5 $\pm$  1.2 &    -2.38 $\pm$  0.05 &  -111.3 $\pm$   0.1 &   1.93~~ &    0.000 $\pm$  0.001 &   -0.000 $\pm$  0.001\\
J1609+2641 &    4.650 $\pm$  0.003 &    -0.430 $\pm$ 0.006 &    -0.446 $\pm$ 0.019 &    0.24 $\pm$  0.005 &   -52.3 $\pm$  0.7 &    -0.71 $\pm$  0.14 &     3.8 $\pm$   0.2 &   0.24~~ &    0.069 $\pm$  0.001 &    0.065 $\pm$  0.002\\
J1619-8418 &    1.513 $\pm$  0.002 &    -1.057 $\pm$ 0.010 &     0.034 $\pm$ 0.028 &    0.04 $\pm$  0.005 &    40.7 $\pm$ 14.5 &    -0.13 $\pm$  0.18 &     1.9 $\pm$   0.7 &   0.05~~ &    0.018 $\pm$  0.001 &    0.014 $\pm$  0.003\\
J1726-5529 &    5.237 $\pm$  0.003 &    -0.078 $\pm$ 0.005 &    -0.331 $\pm$ 0.018 &    0.13 $\pm$  0.003 &   -71.2 $\pm$  0.6 &     0.43 $\pm$  0.13 &    -1.1 $\pm$   0.2 &   0.18~~ &   -0.014 $\pm$  0.001 &   -0.012 $\pm$  0.001\\
J1733-1304 &    6.217 $\pm$  0.004 &    -0.303 $\pm$ 0.006 &    -0.017 $\pm$ 0.019 &    3.58 $\pm$  0.003 &   -85.1 $\pm$  0.9 &    -0.18 $\pm$  0.03 &   -61.5 $\pm$   0.1 &   3.65~~ &    0.057 $\pm$  0.001 &    0.055 $\pm$  0.002\\
J1744-5144 &    6.932 $\pm$  0.003 &    -0.294 $\pm$ 0.004 &    -0.131 $\pm$ 0.013 &    0.10 $\pm$  0.003 &   -14.4 $\pm$  1.6 &    -1.18 $\pm$  0.21 &   -11.3 $\pm$   0.3 &   0.07~~ &   -0.005 $\pm$  0.001 &   -0.006 $\pm$  0.001\\
J1830-3602 &    7.236 $\pm$  0.005 &    -1.305 $\pm$ 0.006 &    -0.075 $\pm$ 0.018 &    0.30 $\pm$  0.002 &   -86.1 $\pm$  0.6 &     0.02 $\pm$  0.13 &    -0.8 $\pm$   0.1 &   0.36~~ &   -0.018 $\pm$  0.001 &   -0.015 $\pm$  0.001\\
J1833-2103 &   10.610 $\pm$  0.006 &     0.063 $\pm$ 0.005 &     0.311 $\pm$ 0.016 &    0.14 $\pm$  0.003 &   -30.4 $\pm$  1.6 &    -1.29 $\pm$  0.15 &    -9.6 $\pm$   0.3 &   0.11~~ &   -0.089 $\pm$  0.001 &   -0.100 $\pm$  0.002\\
J1859-6615 &    1.601 $\pm$  0.002 &    -0.956 $\pm$ 0.010 &     0.074 $\pm$ 0.030 &    0.16 $\pm$  0.004 &     3.3 $\pm$  0.8 &    -2.35 $\pm$  0.19 &     6.8 $\pm$   0.3 &   0.08~~ &    0.001 $\pm$  0.001 &    0.001 $\pm$  0.002\\
J1911-2006 &    2.289 $\pm$  0.002 &     0.235 $\pm$ 0.006 &    -0.127 $\pm$ 0.021 &    1.79 $\pm$  0.003 &   -84.0 $\pm$  1.0 &     0.43 $\pm$  0.02 &   -81.7 $\pm$   0.1 &   1.94~~ &   -0.099 $\pm$  0.001 &   -0.100 $\pm$  0.001\\
J1923-2104 &    1.213 $\pm$  0.001 &    -0.144 $\pm$ 0.008 &     0.128 $\pm$ 0.025 &    2.93 $\pm$  0.004 &   -38.7 $\pm$  0.2 &    -0.00 $\pm$  0.03 &     8.5 $\pm$   0.1 &   2.85~~ &    0.065 $\pm$  0.001 &    0.075 $\pm$  0.002\\
J1924-2914 &    4.908 $\pm$  0.003 &    -0.349 $\pm$ 0.005 &     0.371 $\pm$ 0.017 &    1.31 $\pm$  0.003 &    -2.3 $\pm$  0.2 &    -0.26 $\pm$  0.02 &   -19.9 $\pm$   0.1 &   1.21~~ &   -0.021 $\pm$  0.001 &   -0.005 $\pm$  0.002\\
J1939-6342 &   14.465 $\pm$  0.003 &    -0.266 $\pm$ 0.002 &    -0.673 $\pm$ 0.005 &    0.15 $\pm$  0.001 &    28.1 $\pm$  0.6 &     0.58 $\pm$  0.08 &    -1.1 $\pm$   0.1 &   0.17~~ &    0.043 $\pm$  0.001 &    0.057 $\pm$  0.001\\
J1939-6342 &   14.702 $\pm$  0.005 &    -0.220 $\pm$ 0.003 &    -0.661 $\pm$ 0.010 &    0.16 $\pm$  0.002 &    -6.1 $\pm$  1.5 &    -0.55 $\pm$  0.10 &    -2.7 $\pm$   0.2 &   0.15~~ &    0.016 $\pm$  0.001 &    0.032 $\pm$  0.001\\
J1951-2737 &    1.297 $\pm$  0.002 &    -1.018 $\pm$ 0.011 &     0.014 $\pm$ 0.032 &    1.40 $\pm$  0.005 &   -85.9 $\pm$  0.4 &    -4.54 $\pm$  0.11 &    -2.5 $\pm$   0.1 &   0.87~~ &   -0.007 $\pm$  0.001 &   -0.003 $\pm$  0.002\\
J2007-1016 &    1.502 $\pm$  0.001 &    -0.783 $\pm$ 0.008 &     0.092 $\pm$ 0.023 &    5.18 $\pm$  0.004 &   -18.0 $\pm$  0.9 &    -0.10 $\pm$  0.01 &   -82.8 $\pm$   0.1 &   5.16~~ &    0.005 $\pm$  0.001 &    0.018 $\pm$  0.005\\
J2011-0644 &    2.642 $\pm$  0.002 &    -0.237 $\pm$ 0.007 &    -0.325 $\pm$ 0.024 &    0.25 $\pm$  0.003 &   -37.6 $\pm$  0.3 &     0.08 $\pm$  0.06 &    -0.3 $\pm$   0.2 &   0.25~~ &   -0.065 $\pm$  0.001 &   -0.064 $\pm$  0.001\\
J2052-3640 &    1.366 $\pm$  0.001 &    -1.256 $\pm$ 0.008 &     0.008 $\pm$ 0.024 &    0.19 $\pm$  0.004 &   -85.3 $\pm$  0.7 &     0.04 $\pm$  0.14 &     1.1 $\pm$   0.1 &   0.27~~ &    0.009 $\pm$  0.001 &    0.010 $\pm$  0.002\\
J2130+0502 &    4.033 $\pm$  0.004 &    -0.441 $\pm$ 0.009 &     0.003 $\pm$ 0.028 &    0.07 $\pm$  0.004 &    15.1 $\pm$  2.0 &     0.34 $\pm$  0.20 &     5.7 $\pm$   0.7 &   0.07~~ &    0.028 $\pm$  0.001 &    0.019 $\pm$  0.001\\
J2131-1207 &    1.970 $\pm$  0.002 &     0.108 $\pm$ 0.007 &    -0.010 $\pm$ 0.026 &    1.75 $\pm$  0.004 &   -49.9 $\pm$  0.1 &     0.25 $\pm$  0.03 &     5.5 $\pm$   0.1 &   1.86~~ &   -0.089 $\pm$  0.001 &   -0.076 $\pm$  0.002\\
J2131-2036 &    1.960 $\pm$  0.002 &    -1.018 $\pm$ 0.007 &     0.034 $\pm$ 0.022 &    0.05 $\pm$  0.004 &    51.1 $\pm$ 14.9 &    -0.93 $\pm$  0.30 &    18.0 $\pm$   0.6 &   0.05~~ &    0.036 $\pm$  0.001 &    0.027 $\pm$  0.002\\
J2134-0153 &    1.843 $\pm$  0.001 &     0.098 $\pm$ 0.007 &     0.299 $\pm$ 0.022 &    3.91 $\pm$  0.004 &    42.3 $\pm$  0.6 &    -0.94 $\pm$  0.02 &    47.4 $\pm$   0.1 &   3.57~~ &   -0.040 $\pm$  0.001 &   -0.061 $\pm$  0.003\\
J2136+0041 &    3.849 $\pm$  0.004 &     1.473 $\pm$ 0.007 &     0.291 $\pm$ 0.029 &    0.33 $\pm$  0.004 &    -7.2 $\pm$  0.6 &     0.86 $\pm$  0.11 &    -1.3 $\pm$   0.2 &   0.44~~ &   -0.090 $\pm$  0.001 &   -0.112 $\pm$  0.010\\
J2148+0657 &    3.267 $\pm$  0.003 &     0.044 $\pm$ 0.007 &     0.128 $\pm$ 0.025 &    0.72 $\pm$  0.004 &   -17.8 $\pm$  0.4 &    -1.21 $\pm$  0.15 &     5.8 $\pm$   0.1 &   0.72~~ &   -0.225 $\pm$  0.001 &   -0.213 $\pm$  0.002\\
J2152-2828 &    2.917 $\pm$  0.003 &    -0.656 $\pm$ 0.008 &     0.024 $\pm$ 0.025 &    2.89 $\pm$  0.004 &    31.1 $\pm$  0.6 &    -2.99 $\pm$  0.06 &   -41.7 $\pm$   0.1 &   2.15~~ &   -0.005 $\pm$  0.001 &   -0.011 $\pm$  0.009\\
J2158-1501 &    4.055 $\pm$  0.002 &    -0.119 $\pm$ 0.005 &    -0.034 $\pm$ 0.017 &    4.61 $\pm$  0.003 &   -58.6 $\pm$  0.2 &    -0.24 $\pm$  0.03 &    13.6 $\pm$   0.1 &   4.45~~ &   -0.100 $\pm$  0.001 &   -0.102 $\pm$  0.002\\
J2206-1835 &    6.273 $\pm$  0.004 &    -0.315 $\pm$ 0.005 &     0.107 $\pm$ 0.017 &    0.15 $\pm$  0.003 &   -48.3 $\pm$  1.2 &     0.52 $\pm$  0.09 &    15.4 $\pm$   0.2 &   0.15~~ &   -0.009 $\pm$  0.001 &   -0.007 $\pm$  0.001\\
J2212+0152 &    2.917 $\pm$  0.003 &    -0.596 $\pm$ 0.008 &    -0.049 $\pm$ 0.026 &    0.49 $\pm$  0.004 &   -11.5 $\pm$  0.5 &    -0.29 $\pm$  0.10 &     0.8 $\pm$   0.1 &   0.57~~ &    0.017 $\pm$  0.001 &    0.017 $\pm$  0.002\\
J2214-3835 &    1.808 $\pm$  0.002 &    -0.561 $\pm$ 0.008 &     0.066 $\pm$ 0.025 &    0.57 $\pm$  0.004 &   -84.7 $\pm$  0.5 &    -0.32 $\pm$  0.11 &     0.8 $\pm$   0.1 &   0.62~~ &   -0.000 $\pm$  0.001 &   -0.000 $\pm$  0.001\\
J2225-0457 &    7.708 $\pm$  0.004 &    -0.442 $\pm$ 0.005 &     0.072 $\pm$ 0.016 &    3.89 $\pm$  0.003 &   -66.6 $\pm$  0.4 &    -0.53 $\pm$  0.01 &   -28.6 $\pm$   0.1 &   3.63~~ &   -0.033 $\pm$  0.001 &   -0.027 $\pm$  0.002\\
J2229-3823 &    2.034 $\pm$  0.002 &    -1.263 $\pm$ 0.008 &    -0.009 $\pm$ 0.023 &    0.46 $\pm$  0.004 &    51.8 $\pm$ 18.8 &    -0.58 $\pm$  0.11 &     5.1 $\pm$   0.1 &   0.48~~ &    0.001 $\pm$  0.001 &   -0.000 $\pm$  0.001\\
J2232+1143 &    6.940 $\pm$  0.004 &    -0.399 $\pm$ 0.005 &    -0.014 $\pm$ 0.017 &    3.02 $\pm$  0.003 &   -42.1 $\pm$ 19.5 &    -0.06 $\pm$  0.07 &   -54.5 $\pm$   0.1 &   3.41~~ &   -0.004 $\pm$  0.001 &   -0.002 $\pm$  0.003\\
J2236+2828 &    1.789 $\pm$  0.002 &     0.324 $\pm$ 0.009 &     0.203 $\pm$ 0.033 &    0.46 $\pm$  0.005 &    11.5 $\pm$ 21.7 &    -0.74 $\pm$  0.25 &  -126.6 $\pm$   0.1 &   0.56~~ &   -0.042 $\pm$  0.001 &   -0.023 $\pm$  0.003\\
J2246-1206 &    1.781 $\pm$  0.002 &     0.405 $\pm$ 0.009 &     0.347 $\pm$ 0.031 &    1.56 $\pm$  0.004 &    66.7 $\pm$  0.3 &     0.05 $\pm$  0.04 &   -17.5 $\pm$   0.1 &   1.56~~ &   -0.102 $\pm$  0.001 &   -0.091 $\pm$  0.002\\
J2253+1608 &   16.137 $\pm$  0.008 &    -0.123 $\pm$ 0.004 &     0.264 $\pm$ 0.014 &    6.11 $\pm$  0.006 &    57.5 $\pm$  0.6 &    -0.49 $\pm$  0.02 &   -56.2 $\pm$   0.1 &   5.60~~ &    0.126 $\pm$  0.001 &    0.115 $\pm$  0.006\\

\enddata

\end{deluxetable}
\end{longrotatetable}




\end{document}